\documentclass[%
superscriptaddress,
 floatfix,
 amsmath,amssymb,
 aps, prx, 
  reprint
]{revtex4-2}
\usepackage[normalem]{ulem}
\usepackage[dvipsnames]{xcolor}
\usepackage{graphicx}
\usepackage{dcolumn}
\usepackage{bm}
\usepackage[dvipsnames]{xcolor}
\usepackage{amsmath}
\makeatletter
\let\save@mathaccent\mathaccent
\newcommand*\if@single[3]{%
  \setbox0\hbox{${\mathaccent"0362{#1}}^H$}%
  \setbox2\hbox{${\mathaccent"0362{\kern0pt#1}}^H$}%
  \ifdim\ht0=\ht2 #3\else #2\fi
  }
\newcommand*\rel@kern[1]{\kern#1\dimexpr\macc@kerna}
\newcommand*\widebar[1]{\@ifnextchar^{{\wide@bar{#1}{0}}}{\wide@bar{#1}{1}}}
\newcommand*\wide@bar[2]{\if@single{#1}{\wide@bar@{#1}{#2}{1}}{\wide@bar@{#1}{#2}{2}}}
\newcommand*\wide@bar@[3]{%
  \begingroup
  \def\mathaccent##1##2{%
    \let\mathaccent\save@mathaccent
    \if#32 \let\macc@nucleus\first@char \fi
    \setbox\z@\hbox{$\macc@style{\macc@nucleus}_{}$}%
    \setbox\tw@\hbox{$\macc@style{\macc@nucleus}{}_{}$}%
    \dimen@\wd\tw@
    \advance\dimen@-\wd\z@
    \divide\dimen@ 3
    \@tempdima\wd\tw@
    \advance\@tempdima-\scriptspace
    \divide\@tempdima 10
    \advance\dimen@-\@tempdima
    \ifdim\dimen@>\z@ \dimen@0pt\fi
    \rel@kern{0.6}\kern-\dimen@
    \if#31
      \overline{\rel@kern{-0.6}\kern\dimen@\macc@nucleus\rel@kern{0.4}\kern\dimen@}%
      \advance\dimen@0.4\dimexpr\macc@kerna
      \let\final@kern#2%
      \ifdim\dimen@<\z@ \let\final@kern1\fi
      \if\final@kern1 \kern-\dimen@\fi
    \else
      \overline{\rel@kern{-0.6}\kern\dimen@#1}%
    \fi
  }%
  \macc@depth\@ne
  \let\math@bgroup\@empty \let\math@egroup\macc@set@skewchar
  \mathsurround\z@ \frozen@everymath{\mathgroup\macc@group\relax}%
  \macc@set@skewchar\relax
  \let\mathaccentV\macc@nested@a
  \if#31
    \macc@nested@a\relax111{#1}%
  \else
    \def\gobble@till@marker##1\endmarker{}%
    \futurelet\first@char\gobble@till@marker#1\endmarker
    \ifcat\noexpand\first@char A\else
      \def\first@char{}%
    \fi
    \macc@nested@a\relax111{\first@char}%
  \fi
  \endgroup
}
\makeatother
\usepackage{braket}
\usepackage{appendix}
\usepackage{hyperref}
\usepackage{tabularx}
\usepackage{multirow}
\usepackage{enumitem}
\usepackage{slashbox}
\usepackage{comment}

\newcommand{\ofr}{(\mathbf{r})}
\newcommand{\mk}{\mathbf{q}}
 
\begin{document}


\title{Neutron scattering from local magnetoelectric multipoles: a combined theoretical, computational, and experimental perspective}

\author{Andrea Urru}
\affiliation{Materials Theory, ETH Zurich, Wolfgang-Pauli-Strasse 27, 8093 Zurich, Switzerland} 
\author{Jian-Rui Soh}%
\affiliation{Institute  of  Physics,  \'Ecole  Polytechnique  F\'ed\'erale  de  Lausanne  (EPFL),  CH-1015  Lausanne,  Switzerland}%
\author{Navid Qureshi}
\affiliation{Institut Laue-Langevin, 6 rue Jules Horowitz, BP 156, 38042 Grenoble Cedex 9, France}%
\author{Anne Stunault}
\affiliation{Institut Laue-Langevin, 6 rue Jules Horowitz, BP 156, 38042 Grenoble Cedex 9, France}%
\author{Bertrand Roessli}
\affiliation{Laboratory for Neutron Scattering and Imaging, Paul Scherrer Institut, Villigen, Switzerland}%
\author{Henrik M. Rønnow}
\affiliation{Institute  of  Physics,  \'Ecole  Polytechnique  F\'ed\'erale  de  Lausanne  (EPFL),  CH-1015  Lausanne,  Switzerland}%
\author{Nicola A. Spaldin}
\affiliation{Materials Theory, ETH Zurich, Wolfgang-Pauli-Strasse 27, 8093 Zurich, Switzerland}

\date{\today}

\begin{abstract}
We address magnetic neutron scattering in the presence of local non-centrosymmetric asymmetries of the magnetization density. Such inversion-symmetry breaking, combined with the absence of time-reversal symmetry, can be described in terms of magnetoelectric multipoles which form the second term after the magnetic dipole in the multipole expansion of the magnetization density. We provide a pedagogical review of the theoretical formalism of magnetic neutron diffraction in terms of the multipole expansion of the scattering cross-section. In particular, we show how to compute the contribution of magnetoelectric multipoles to the scattering amplitude starting from \textit{ab initio} calculations. We also provide general guidelines on how to experimentally detect long-ranged order of magnetoelectric multipoles using either unpolarized or polarized neutron scattering.  
As a case study, we search for the presence of magnetoelectric multipoles in CuO by comparing theoretical first-principle predictions with experimental spherical neutron polarimetry measurements. 

\end{abstract}

\maketitle

\section{Introduction} 
\label{Intro}
Many macroscopic physical properties of crystalline solids are governed by the underlying electronic charge ($n (\mathbf{r})$) and magnetization ($\boldsymbol{\mu} (\mathbf{r})$) densities. Hence, to understand the behaviour of materials, it is imperative to characterize these densities accurately. While core electrons fill closed shells with spherically symmetric charge densities and no magnetic moment, the more delocalized valence electrons can have uncompensated magnetic moments and asymmetric charge densities. These in turn affect the magnetic and dielectric properties. When time-reversal and spatial-inversion symmetries are both broken, the magnetic ($\boldsymbol{\mu}(\mathbf{r})$) and electric (\textbf{r}) degrees of freedom become intertwined, giving rise to a non-centrosymmetric valence magnetization density, for which $\boldsymbol{\mu}(\mathbf{r})\otimes\mathbf{r}\neq0$, which is the central focus of this work.

Neutrons can probe the magnetic field distributions arising from these magnetization densities~\cite{Boothroyd_book}. 
In this work we focus on the interaction between neutrons and the magnetic field generated by electrons in crystalline solids. The cross-section of this process is small and therefore it is computed using the Born approximation and Fermi's golden rule, which leads to an expression in terms of the transverse magnetization, $\mathbf{M}_{\perp} (\mathbf{q})$ \cite{Boothroyd_book}, with $\mathbf{q}$ being the momentum transferred in the process. $\mathbf{M}_{\perp} (\mathbf{q})$ is often expanded in spherical harmonics and typically only the leading-order terms, proportional to the spin and orbital angular momenta, are taken into account. However, as these terms do not capture the non-centrosymmetric asymmetries in the magnetization density, a complete description should include higher-order multipolar terms in the expansion of $\mathbf{M}_{\perp} (\mathbf{q})$. 

The first terms beyond the dipole contribution contain multipoles that break both time-reversal and spatial-inversion symmetries and are thus essential for describing the non-centrosymmetric asymmetries of the magnetization density. These multipoles have been demonstrated to be intimately connected to the linear magnetoelectric (ME) effect \cite{me_multipoles}, and as such they are called \textit{ME multipoles}. In their irreducible form, they are the scalar ME monopole 
\begin{equation}
   a = \frac{1}{3} \int \mathbf{r} \cdot  \boldsymbol{\mu}(\mathbf{r}) \, d^3 \mathbf{r} ,
\end{equation}
the ME toroidal moment vector 
\begin{equation}
    \mathbf{t} = \frac{1}{2}  \int \mathbf{r} \times \boldsymbol{\mu}(\mathbf{r}) \, d^3 \mathbf{r},
\end{equation}
and the ME quadrupole tensor 
\begin{equation}
Q_{ij} = \frac{1}{2} \int \left[r_i \mu_j (\mathbf{r}) + r_j \mu_i (\mathbf{r}) - \frac{2}{3} \delta_{ij} \mathbf{r}\! \cdot \boldsymbol{\mu}(\mathbf{r}) \right] d^3\mathbf{r},
\end{equation}
which correspond respectively to the trace, the anti-symmetric part, and the symmetric traceless part of the so-called ME multipole tensor \cite{me_multipoles}, defined as 
\begin{equation}
    \mathcal{M}_{ij} = \int r_{i} {\mu_j(\mathbf{r})} d^3 \mathbf{r}.
\end{equation}

The expansion of the spin and orbital contributions to the magnetic neutron scattering in terms of irreducible multipole tensors was first discussed in general terms in the 1960s by M. Blume~\cite{Blume}, D. F. Johnston \cite{Johnston_orbital} and S. W. Lovesey \cite{Lovesey_review_multipoles, Lovesey_multipoles_note, Lovesey_book_right}. Renewed interest in ME multiferroics and the linear ME effect, as well as a report of orbital currents in CuO that were interpreted in terms of toroidal moments \cite{toroidal_CuO}, motivated recent specific analysis of the inversion-breaking ME multipoles, with particular emphasis on the toroidal moment $\mathbf{t}$ \cite{Lovesey_multipoles_1, Lovesey_multipoles_2}. Several case studies for specific materials \cite{Lovesey_multipoles_4, Lovesey_multipoles_3, Lovesey_multipoles_5, Lovesey_multipoles_pnictides, Lovesey_anapoles} followed, in which the multipolar contributions for the magnetic ions were computed using a phenomenological method based on linear combinations of atomic orbitals.

Contemporary \textit{ab initio} density functional theory (DFT) \cite{DFT_HK, DFT_KS} in its non-collinear formulation provides an accurate description of the magnetization density, including its asymmetries dictated by the local atomic environment.
Access to the full $\boldsymbol{\mu} (\mathbf{r})$ makes DFT ideally suited for the computation of the aforementioned ME multipoles. A convenient technique implemented in the \texttt{ELK} open-source package \cite{Elk_code} exists for extracting these ME multipoles from the density matrix calculated in a Hubbard $U$-based formalism \cite{multipole_decomposition, Liechtenstein_U, Dudarev_U, U_cococcioni}. Yet, the connection between the ME multipoles and the experimentally accessible neutron scattering cross section, is still lacking. 

In this work, we build on these earlier developments, and show how to bridge first-principles calculations of the size of the ME multipoles to the theoretical formulation of elastic magnetic neutron scattering to give a quantitative interpretation of experimental data. Our goal is to provide the tools needed to assess possible signatures of ME multipoles in a neutron diffraction experiment: to this end, we take added care to cast the ME diffraction formalism into the terminology widely used in conventional magnetic neutron scattering. We use our combined approach to address the existence of ME multipoles in CuO. Using DFT, we predict that CuO displays an anti-ferroic arrangement of ME multipoles, commensurate with the magnetic order. Remarkably, we demonstrate that the largest multipoles reside on the formally non-magnetic oxygen ligands, which have been neglected in earlier neutron scattering analyses \cite{Lovesey_multipoles_1}. In light of our first-principles predictions, we analyze our spherical neutron polarimetry (SNP) data of CuO. We find that the experimental results are consistent with the presence of ME multipoles, but additional, more accurate data on the off-diagonal entries of the polarization matrix are required to make a decisive, unambiguous proof of the observation of such hidden multipolar order.

The paper is organized in the following way. In Section \ref{neutron_theory} we review the theoretical approach developed by Lovesey \cite{Lovesey_book_right, Lovesey_multipoles_1}, connect it to the \textit{ab initio} calculation of multipoles, and provide the reader with general guidelines on how to detect hidden ME multipolar orders in conventional neutron scattering and spherical neutron polarimetry. In Section \ref{CuO} we discuss in detail our DFT predictions and experimental results on CuO. Finally, Section \ref{final} contains a summary of our findings and the future perspectives of the present work.

\section{Neutron cross-section for ME multipoles} 
\label{neutron_theory}

The main experimentally accessible quantity in a neutron scattering experiment is the intensity of scattered neutrons $I(\mathbf{q})$, which depends on the momentum $\mathbf{q}$ transferred in the scattering process. $I(\mathbf{q})$ is the \textit{partial differential cross-section}, 
\begin{equation}
\label{eq_cross_sec}
I(\mathbf{q}) = \frac{d^2 \sigma}{d \Omega d E_f},
\end{equation}
defined in terms of the total scattering cross section $\sigma$, as the number of neutrons scattered per unit of time into solid angle $d \Omega$ about a given direction, with final energy in the interval $(E_f, E_f + \delta E_f)$. Following Fermi's Golden Rule, the cross section is proportional to the Fourier component at $\mathbf{q}$ of the interaction potential $V(\mathbf{r})$ between the neutrons and the target: 
\begin{equation}
\label{eq_int}
I(\mathbf{q}) \propto \lvert V(\mathbf{q}) \lvert ^2.
\end{equation}
Here we consider the interaction between the neutron spin angular momentum $\gamma \mu_{\text{N}} \boldsymbol{\sigma}$, with $\gamma$ the neutron gyromagnetic ratio, and the magnetic field $\mathbf{H} (\mathbf{r})$ generated by the electrons through their spin and orbital angular momentum. The interaction potential reads 
\begin{equation}
V \ofr = - \gamma \mu_{\text{N}} \bm{\sigma} \cdot \mathbf{H}\ofr,
\end{equation}
and its Fourier component at $\mathbf{q}$ is usually written as: 
\begin{equation}
\label{eq17}
V(\mathbf{q}) = \gamma \mu_0 \mu_{\text{N}} \boldsymbol{\sigma} \cdot \mathbf{M}_{\perp} (\mathbf{q}).
\end{equation}
$\mathbf{M}_{\perp} (\mathbf{q})$ is called the \textit{transverse magnetization} and is computed as \cite{Boothroyd_book}
\begin{equation}
\label{eq1}
    \mathbf{M}_{\perp} (\mk) = \mk \wedge (\mathbf{M}(\mk) \wedge \mk),  
\end{equation}
where $\mathbf{M}(\mk)$ is the Fourier transform of the magnetization density. Eqs. \eqref{eq17} and \eqref{eq1} imply that neutrons are only sensitive to the component of the magnetization perpendicular to the scattering vector $\mathbf{q}$.

In a solid, $\mathbf{M}(\mk)$ reads 
\begin{equation}
\label{eq2}
    \mathbf{M} (\mk) = \sum_j e^{i \mk \cdot \mathbf{R}_j } e^{i \mk \cdot \mathbf{r}_j} \, \left( \mathbf{s}_j - \frac{\hat{\mk} \wedge \boldsymbol{\nabla}_j}{q} \right).
\end{equation}
Here, the index $j$ labels the unpaired electrons of the atoms in the solid, $\mathbf{R}_j$ is the position of the ion that the $j$-th electron belongs to, and $\mathbf{r}_j$ and $\mathbf{s}_j$ are the position and spin operators of the $j$-th electron, respectively. $\boldsymbol{\nabla}_j$ is related to the momentum operator of the $j$-th electron via $\mathbf{p}_j = - i \hbar \boldsymbol{\nabla}_j$. This expression, $\mathbf{M} (\mk)$, fully accounts for the interaction between neutrons and the spin and orbital magnetization density created by the electrons in a solid. As mentioned previously, however, the higher-order terms in the expansion of $\mathbf{M} (\mk)$ beyond the magnetic dipole approximation described by the first term, are usually neglected. In particular, the ME multipolar terms in $\mathbf{M} (\mk)$ are usually omitted. In the sections that follow, we consider the contribution of the ME multipoles to $\mathbf{M} (\mk)$ from the theoretical, computational, and experimental perspectives. In section~\ref{theory}, we outline which term in the expansion of the expression for $\mathbf{M}(\mk)$ constitutes the ME multipoles with particular emphasis on how they contribute to the neutron scattering cross section. Next, in section~\ref{multipole_calc}, we outline how first-principles calculations can be used to obtain $\mathbf{M}(\mk)$, including the components from ME multipoles. Finally, in section~\ref{Experiment}, we discuss how ME multipoles affect the polarization of neutrons, since this is the main quantity measured in spherical neutron polarimetry experiments.

\subsection{\texorpdfstring{Contribution of ME multipoles to $\mathbf{M}(\mathbf{q})$: a theoretical perspective}{me theo}}
\label{theory}
\subsubsection{Spin contribution}
\label{spin_contribution}
\paragraph{Multipole expansion.}

In the following, we focus on the spin term of Eq. \eqref{eq2} for a single unpaired electron. We identify with $s_p$ the $p$-th spherical component of $\mathbf{s}$ and expand $\exp(i \mathbf{q} \cdot \mathbf{r})$ in spherical harmonics $Y_{LM}$, to obtain \cite{Lovesey_book_right, Lovesey_multipoles_1}: 
\begin{equation}
\label{eq3}
\begin{split}
     e^{i \mk \cdot \mathbf{r}} s_p & = 4 \pi \sum_{L,L'} (-i)^{L} \sqrt{\frac{2 L'+1}{3}} j_L(qr) \sum_{M, M'} Y_{LM} (\hat{\mk}) \\ & \times \left[ \sum_{\bar{p}, \widebar{M}} s_{\bar{p}} Y_{L \widebar{M}} (\hat{\mathbf{r}}) \langle 1 \bar{p} \, L \widebar{M} \lvert L' M' \rangle \right] \langle L M \, L' M' \lvert 1 p \rangle, 
\end{split}
\end{equation}
where $j_L(qr)$ is the spherical Bessel function of order $L$, and $\langle 1 \bar{p} \, L \widebar{M} \lvert L' M' \rangle$ and $\langle L M \, L' M' \lvert 1 p \rangle$ are Clebsch-Gordan coefficients. The quantity inside the square brackets in Eq. \eqref{eq3} corresponds to the tensor product of a spherical tensor of rank 1 in the spin variables with a spherical tensor of rank $L$ in the spatial variables: the result is an irreducible spherical tensor of rank $L'$, that we identify as: 
\begin{equation}
\label{eq4}
\begin{split}
    T^{(\text{s})}_{L' M'} & = \sqrt{4 \pi} \left[ \sum_{\bar{p} \widebar{M}} s_{\bar{p}} Y_{L \widebar{M}} (\hat{\mathbf{r}}) \langle 1 \bar{p} \, L \widebar{M} \lvert L' M' \rangle \right] \\ 
    & = \sqrt{4 \pi} \left[ \mathbf{s} \otimes \mathbf{Y}_L (\hat{\mathbf{r}}) \right]^{L'}_{M'}.
\end{split}
\end{equation}
The rank $L'$ of the resulting tensor is related to $L$ following the composition rules, $\lvert L - 1 \lvert \le L' \le L + 1$, and the spherical component $M'$ runs from $-L'$ to $L'$ in integer steps. $\mathbf{T}^{(\text{s})}_{L'}$ always breaks time-reversal symmetry, and for odd $L$ it breaks inversion symmetry as well. Using the definition of $T^{(\text{s})}_{L' M'}$, the sum over $M$ and $M'$ in Eq. \eqref{eq3} can be written as a tensor product itself: 
\begin{equation}
\begin{split}
\label{eq5}
e^{i \mk \cdot \mathbf{r}} s_p = &\sum_{L, L'} (-i)^{L} \sqrt{\frac{2 L'+1}{3}} j_L(qr) \\ 
&\times \left\{ \sqrt{4 \pi} \left[ \mathbf{Y}_{L} (\hat{\mk}) \otimes \mathbf{T}^{(\text{s})}_{L'} \right]^{1}_{p} \right\}.
\end{split}
\end{equation}
The tensor product $\mathbf{Y}_{L} (\hat{\mk}) \otimes \mathbf{T}^{(\text{s})}_{L'}$ identifies how the multipole tensor operators couple to the scattering wave vector to produce a vector in the magnetization $\mathbf{M} (\mk)$. 

In a practical calculation, we are interested in computing the expectation value of $e^{i \mk \cdot \mathbf{r}} \mathbf{s}$. We describe the unpaired electron with a wave function $\Psi_{n l} (\mathbf{r}, s)$, which we factorize into a radial part $R_{n l} (r)$, an angular part written in the spherical harmonics basis set, and a spin part $\chi (s)$ as: 
\begin{equation}
    \Psi_{n l} (\mathbf{r}, s) = \sum_{\sigma} \sum_{m_l} \alpha^{\sigma}_{n l m_l} R_{n l} (r) \, Y_{l \, m_l} (\Omega) \, \chi_{\frac{1}{2} \, \sigma} (s),
\end{equation}
where $\sigma=- 1/2, 1/2$. The expectation value of $e^{i \mk \cdot \mathbf{r}} \mathbf{s}$ is then: 
\begin{equation}
\label{eq7}
\begin{split}
\langle e^{i \mk \cdot \mathbf{r}} s_p \rangle = & \sum_{l, l'} \sum_{L, L'} (- i)^L \sqrt{\frac{2 L' + 1}{3}} \, I^1_{L, ll'} (q) \\ 
& \times \bigg \langle \sqrt{4 \pi} \left[ \mathbf{Y}_{L} (\hat{\mk}) \otimes \mathbf{T}^{(\text{s})}_{L'} \right]^{1}_{p} \bigg \rangle_{l l'},
\end{split}
\end{equation}
where $I^1_{L, l l'} (q)$ is the radial integral defined as: 
\begin{equation}
    I^1_{L, l l'} (q)= \int_0^{+\infty} R_{n l} (r) \, j_L (k r) \, R_{n l'} (r) \, r^2 \, dr,
\end{equation}
and the quantity in brackets is the $l l'$ matrix element of the tensor operator $\sqrt{4 \pi} \left[ \mathbf{Y}_{L} (\hat{\mk}) \otimes \mathbf{T}^{(\text{s})}_{L'} \right]^{1}$: 
\begin{equation}
\begin{split}
   & \bigg \langle \sqrt{4 \pi} \left[ \mathbf{Y}_{L} (\hat{\mk}) \otimes \mathbf{T}^{(\text{s})}_{L'} \right]^{1}_{p} \bigg \rangle_{l l'} = \sum_{\sigma, \sigma'} \sum_{m_l, m_l'} \alpha^{\sigma \, *}_{n l m_l} \alpha^{\sigma'}_{n l' m_l'} \\ 
   & \times \sqrt{4 \pi} \bigg \langle l, m_l, \frac{1}{2}, \sigma  \bigg \lvert \left[ \mathbf{Y}_{L} (\hat{\mk}) \otimes \mathbf{T}^{(\text{s})}_{L'} \right]^{1}_{p} \bigg \lvert l', m_l', \frac{1}{2}, \sigma' \bigg \rangle .
   \end{split}
\end{equation}

\begin{figure}[t]
\includegraphics[width=0.49\textwidth]{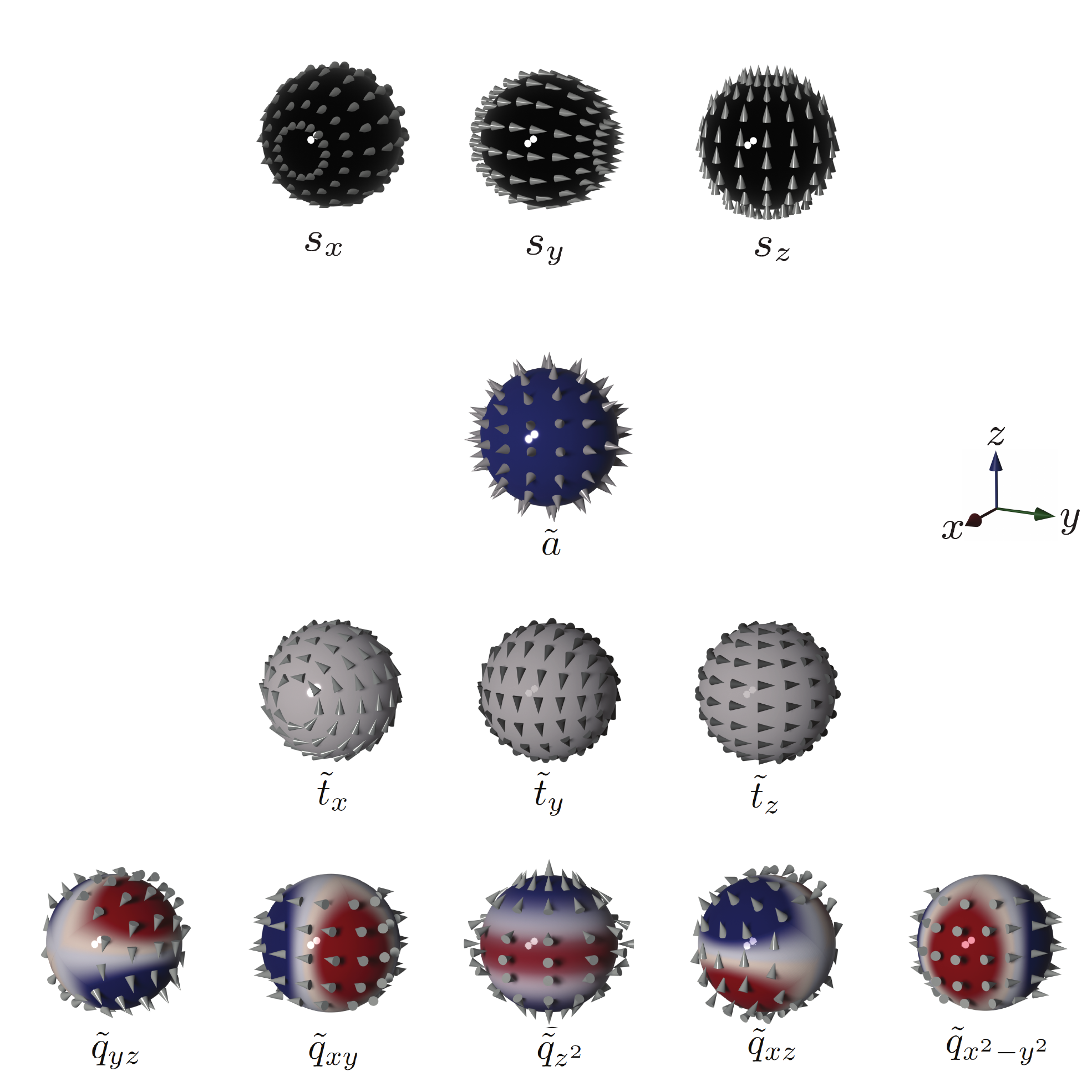}
\caption{\label{fig:MRmultipoles} Representation of the three cartesian components of the magnetic dipole moment $s_x$, $s_y$, $s_z$ (top row). The ME multipoles can be decomposed into three irreducible representations: the pseudo-scalar monopole $\widetilde{a}$ (second row), the anti-symmetric toroidal moment $\widetilde{t}_x$, $\widetilde{t}_y$, $\widetilde{t}_z$ (third row) and the traceless symmetric quadrupoles $\widetilde{q}_{yz}$, $\widetilde{q}_{xy}$, $\widetilde{q}_{z^2}$, $\widetilde{q}_{xz}$, $\widetilde{q}_{x^2-y^2}$ (bottom row). The color schemes of the surfaces of the ME multipoles are defined based on the relative orientation of the position vector \textbf{r} defined with respect to the center of the sphere and the local magnetization density \boldsymbol{$\mu$} (shown with grey arrows), with blue, red and white denoting parallel, anti-parallel and orthogonal \boldsymbol{$\mu$} with respect to \textbf{r} vectors.}
\end{figure}

\paragraph{Leading-order terms.}
Hereafter, we focus on the leading-order terms of the multipole expansion reported in Eq. \eqref{eq5}, i.e. the contributions for $L=0$ and $L=1$. We start from the irreducible spherical tensor operators $T^{(\text{s})}_{L' M'}$, which identify the magnetic multipoles: note that the expression of $T^{(\text{s})}_{L' M'}$ in Eq. \eqref{eq4} contains the unit vector $\hat{\mathbf{r}}$, hence only the angular part of the multipoles, identified with a tilde ($\sim$) in the following, must be considered. 
\begin{itemize}
\item{$L = 0$:} $L'$ can take only the value $1$. Since $Y_{00} (\hat{\mathbf{r}}) = 1 / {\sqrt{4 \pi}}$, $\mathbf{T}^{(\text{s})}_1 = \mathbf{s}$ is the magnetic dipole moment operator;
\item{$L = 1$:} in this case the multipole is linear in both $\hat{\mathbf{r}}$ and $\mathbf{s}$, hence it belongs to the family of ME multipoles, introduced in Ref. \cite{me_multipoles}. $L'$ can take the values $0, 1$, and $2$. For $L' = 0$, the rank-$0$ tensor $T^{(\text{s})}_0$, i.e. a scalar, is proportional to the angular part of the ME monopole operator, $\widetilde{a} = \hat{\mathbf{r}} \cdot \mathbf{s}$. For $L' = 1$, the rank-$1$ tensor $\mathbf{T}^{(\text{s})}_1$ is a vector operator, which corresponds to the angular part of the toroidal moment operator, $\widetilde{\mathbf{t}} = \hat{\mathbf{r}} \wedge \mathbf{s}$. Finally, the case $L' = 2$ corresponds to the angular part of the totally symmetric, traceless, ME quadrupole tensor $\widetilde{Q}$, with entries
    \begin{equation}
        \widetilde{Q}_{ij} = \frac{1}{2} \left[\hat{r}_i s_j +
        \hat{r}_j s_i - \frac{2}{3} \delta_{ij} \hat{\mathbf{r}} \cdot \mathbf{s} \right].
    \end{equation}
In particular, the $5$ components of $\mathbf{T}^{(\text{s})}_2$ are the irreducible spherical components of $\widetilde{Q}$: 
\begin{equation}
    \widetilde{Q} = \begin{pmatrix} \frac{1}{2} \left(\widetilde{q}_{x^2-y^2} - \widetilde{q}_{z^2}\right) & \widetilde{q}_{xy} & \widetilde{q}_{xz} \\ \widetilde{q}_{xy} & - \frac{1}{2} \left(\widetilde{q}_{x^2-y^2}+\widetilde{q}_{z^2} \right) & \widetilde{q}_{yz} \\ \widetilde{q}_{xz} & \widetilde{q}_{yz} & \widetilde{q}_{z^2} \end{pmatrix}.
    \end{equation}
\end{itemize}
A pictorial illustration of these ME multipoles is provided in Fig. \ref{fig:MRmultipoles}.

Higher values of $L$ correspond to higher-order multipoles: as an example, $L=2$ corresponds to magnetic octupoles \cite{mag_oct}, $L=3$ corresponds to magnetic hexadecapoles, and so on.  
The results for $L = 0$ and $L = 1$ are summarized in Table \ref{t1} and a more detailed derivation is provided in Appendix \ref{app_tensor_multipoles}. 
\begingroup
\begin{table}
\setlength{\tabcolsep}{8pt}
\renewcommand{\arraystretch}{1.2}
\begin{tabular}{|c|c|c|c|c|}
\hline
 $L$ & $L'$ & $\mathbf{T}^{(\text{s})}_{L'}$ & Racah's & Spherical \\
 & & & notation & components \\
\hline
0 & 1 & $\mathbf{s}$ & $\mathbf{w}^{011}$ & $s_{-1}$, $s_0$, $s_1$ \\
\hline
\hline
1 & 0 & $\propto \mathbf{\hat{r}} \cdot \mathbf{s}$ & $\mathbf{w}^{110}$ & $\widetilde{a}$ \\
\hline 
1 & 1 & $\propto \mathbf{\hat{r}} \wedge \mathbf{s}$ & $\mathbf{w}^{111}$ & $\widetilde{t}_{-1}$, $\widetilde{t}_0$, $\widetilde{t}_1$ \\ 
\hline 
1 & 2 & $\propto \ket{\hat{\mathbf{r}}} \bra{\mathbf{s}}$ & $\mathbf{w}^{112}$ & $\widetilde{q}_{-2}$, $\widetilde{q}_{-1}$, $\widetilde{q}_0$, $\widetilde{q}_1$, $\widetilde{q}_2$ \\ 
\hline
\end{tabular}
\caption{Tensor $\mathbf{T}^{(\text{s})}_{L'}$ defined by Eq. \eqref{eq4}: cases $L=0$ and $L=1$. The first three columns list $L$, $L'$, and the irreducible spherical tensor $\mathbf{T}^{(\text{s})}_{L'}$, respectively. The fourth column lists $\mathbf{T}^{(\text{s})}_{L'}$ in Racah's notation, and the fifth column lists the spherical components of the tensor operator.}
\label{t1}
\end{table}
\endgroup

After introducing in Eq. \eqref{eq5} the expressions for $\mathbf{T}^{(\text{s})}_{L'}$ and evaluating the tensor product $\mathbf{Y}_{L} (\hat{\mk}) \otimes \mathbf{T}^{(\text{s})}_{L'}$ following the approach discussed in Appendix \ref{app_tensor_multipoles}, the terms $L=0$ and $L=1$ of the expansion of $e^{i \mk \cdot \mathbf{r}} \mathbf{s}$ read:
\begin{equation}
    \begin{split}
    \label{eq6}
        e^{i \mk \cdot \mathbf{r}} \mathbf{s} = & j_0(qr) \, \mathbf{s} + i \, j_1(qr) \left( \widetilde{a} \hat{\mk}+ \frac{3}{2} \widetilde{\mathbf{t}} \wedge \hat{\mk} + 3 \widetilde{Q} {\hat{\mk}} \right) + \dots \, . 
    \end{split}
\end{equation}
It is noteworthy that the term containing $\widetilde{a}$ in the multipole expansion does not contribute to the transverse magnetization $\mathbf{M}_{\perp} (\mk)$ (Eq. \eqref{eq1}), which means that neutrons are not sensitive to the ME monopole. This fact can be explained physically: the neutron spin couples to the magnetic field $\mathbf{B} (\mathbf{r})$ inside the material, which for a given spin texture $\mathbf{s} (\mathbf{r})$ reads 
\begin{equation}
\mathbf{B} (\mathbf{r}) = - 2 \mu_{\text{B}} \frac{\mu_0}{4 \pi} \text{curl} \left( \frac{\mathbf{s} \wedge \mathbf{r}}{r^3} \right).
\end{equation}
Since a purely monopolar spin texture is radial ($\mathbf{s}_{\text{mono}} (\mathbf{r}) = s(r) \hat{\mathbf{r}})$, so $\mathbf{B}_{\text{mono}} \propto \hat{\mathbf{r}} \wedge \mathbf{r} = 0$ and hence it can not be detected by neutrons. On the other hand, the toroidal moment ($\widetilde{\mathbf{t}}$) and the quadrupole ($\widetilde{Q}$) in the spin contribution can deflect neutrons.

\subsubsection{Orbital contribution}
\label{orbital_contribution}
\paragraph{Multipole expansion.} 
In a similar fashion to the spin contribution discussed above, the second term inside the parentheses in Eq. \eqref{eq2} can be expanded in terms of multipoles of the orbital angular momentum $\mathbf{L}$. In particular, we take the expansion of the quantity $e^{i \mathbf{q} \cdot \mathbf{r}} ( \hat{\mathbf{q}} \wedge \boldsymbol{\nabla})$. According to Johnston \cite{Johnston_orbital}, and following the detailed derivation by Lovesey \cite{Lovesey_book_right}, the multipole expansion reads 
\begin{equation}
\label{eq10}
\begin{split}
    e^{i \mathbf{q} \cdot \mathbf{r}} \left( \hat{\mathbf{q}} \wedge \boldsymbol{\nabla} \right)_p &= - \frac{4 \pi}{\sqrt{2}} \sum_{L, L', L''} i^{L+1} j_L (qr) \\ 
    & \times \sqrt{(2L+1)(2L'+1)(2L''+1)} \begin{pmatrix} L & 1 & L' \\ 0 & 0 & 0 \end{pmatrix} \\ & \times \begin{Bmatrix} L & 1 & L' \\ 1 & L'' & 1 \end{Bmatrix} \sum_{M', M''} Y_{L' M'} (\hat{\mathbf{q}}) \left[ \sum_{M \bar{p}} Y_{L M} (\hat{\mathbf{r}}) \right. \\ & \times \nabla_{\bar{p}} \langle L M \, 1 \bar{p} \lvert L'' M'' \rangle \Bigg] \langle L' M' \,  L'' M'' \lvert 1 p \rangle, 
    \end{split}
\end{equation}
where the quantity in curly brackets is a Wigner-6j symbol. The Wigner-3j symbol $\begin{pmatrix} L & 1 & L' \\ 0 & 0 & 0 \end{pmatrix}$ sets the condition $L' = L \pm 1$. The quantity in square brackets corresponds to the tensor product of the rank-$L$ tensor in the spatial variables, $\mathbf{Y}_L (\hat{\mathbf{r}})$, and the rank-1 tensor $\boldsymbol{\nabla}$, which gives the rank-$L''$ tensor: 
\begin{equation}
    \label{eq12}
    T^{\text{(orb.)}}_{L'' M''} = \sqrt{4 \pi} \left[ \mathbf{Y}_L (\hat{\mathbf{r}}) \otimes \boldsymbol{\nabla} \right]^{L''}_{M''}, 
\end{equation}
where $\lvert L - 1 \lvert \le L'' \le L + 1$ following the composition rules. By exploiting the tensor $\mathbf{T}^{\text{(orb.)}}_{L''}$ we have just introduced, the sum over $M'$ and $M''$ in Eq. \eqref{eq10} can be written as a tensor product itself. As such we can recast Eq. \eqref{eq10} compactly, in the following way: 
\begin{equation}
\label{eq11}
\begin{split}
    e^{i \mathbf{q} \cdot \mathbf{r}} \left( \hat{\mathbf{q}} \wedge \boldsymbol{\nabla} \right)_p &= - \frac{1}{\sqrt{2}} \sum_{L, L', L''} i^{L+1} j_L (qr) \\ 
    & \times \sqrt{(2L+1)(2L'+1)(2L''+1)} \begin{pmatrix} L & 1 & L' \\ 0 & 0 & 0 \end{pmatrix} \\ & \times \begin{Bmatrix} L & 1 & L' \\ 1 & L'' & 1 \end{Bmatrix} \left\{ \sqrt{4 \pi} \left[ \mathbf{Y}_{L'} (\hat{\mathbf{q}}) \otimes \mathbf{T}^{\text{(orb.)}}_{L''} \right]^1_{p} \right\}
\end{split}
\end{equation}

\paragraph{Leading-order terms.}

In the following we discuss the first two leading terms of the multipole expansion reported in Eq. \eqref{eq11}, corresponding to the cases $L = 0$ and $L = 1$: 
\begingroup
\begin{table}[t]
\begin{tabular}{|c|c|c|c|}
    \hline
    \backslashbox{$L'$}{$L''$} & $0$ & $1$ & $2$ \\ \hline
                $0$             &   $0$   &  $-1/3$ & $0$   \\ \hline
                $2$             &   $0$   &  $1/6$ & $- 1/2 \sqrt{5}$   \\ \hline
\end{tabular} 
\caption{Values of the Wigner-6j coefficient $\begin{Bmatrix} L & 1 & L' \\ 1 & L'' & 1 \end{Bmatrix}$ for the case $L = 1$.}
\label{t4}
\end{table}
\endgroup

\begin{itemize}
    \item $L = 0$: From the constraints on $L'$ and $L''$, it follows that $L'' = L' = 1$. Moreover, $L = 0$ implies $M = 0$, hence $Y_{00} (\hat{\mathbf{r}}) = 1 / \sqrt{4 \pi}$ in Eq. \eqref{eq12}, and $\mathbf{T}^{\text{(orb.)}}_0 = \boldsymbol{\nabla}$. As a consequence, the leading order term reads 
    \begin{equation}
    \label{eq13}
        e^{i \mathbf{q} \cdot \mathbf{r}} \left( \frac{\hat{\mathbf{q}} \wedge \boldsymbol{\nabla}}{q} \right) \Bigg \lvert_{\substack{L = 0 \\ L' = 1 \\ L'' = 1}} = \frac{1}{2} \frac{j_0 (qr)}{q} \hat{\mathbf{q}} \wedge \boldsymbol{\nabla}. 
    \end{equation}
    \item $L = 1$: The possible values for $L'$ and $L''$ are $L' = \{0, 2\}$ and $L'' = \{0, 1, 2\}$, which results in six possible cases. Three of them are suppressed by a vanishing Wigner-6j symbol, as summarized in Table \ref{t4}, and among the remaining three, the only one that is isotropic in reciprocal space and contributes to the dipole approximation, is $L' = 0$, $L'' = 1$: in this case, $\mathbf{T}^{\text{(orb.)}}_{1} \propto \hat{\mathbf{r}} \wedge \boldsymbol{\nabla}$, which is proportional to the orbital angular momentum $\mathbf{L}$. By substitution into Eq. \eqref{eq11}, the dipole contribution of the orbital angular momentum to the neutron magnetic scattering amplitude reads
    \begin{equation}
       e^{i \mathbf{q} \cdot \mathbf{r}} \left( \frac{\hat{\mathbf{q}} \wedge \boldsymbol{\nabla}}{q} \right) \Bigg \lvert_{\substack{L = 1 \\ L' = 0 \\ L'' = 1}} = \frac{1}{2} \left[ j_0 (qr) + j_2 (qr) \right] \mathbf{L}. 
    \end{equation}
\end{itemize}

The expectation value of the orbital part of $\mathbf{M} (\mk)$ depends on the matrix elements, in the spherical harmonics basis set, of the tensor operators in curly brackets in Eq. \eqref{eq11}, similar to the case discussed in Section \ref{spin_contribution} for the spin part. When computing these matrix elements, the contribution obtained for $L = 0$, reported in Eq. \eqref{eq13}, can be further manipulated and written in terms of the matrix element of the unit position operator $\hat{\mathbf{r}}$ and of the orbital toroidal moment $\boldsymbol{\Omega} = \mathbf{L} \wedge \hat{\mathbf{r}} - \hat{\mathbf{r}} \wedge \mathbf{L}$. We refer the interested reader to Refs. \cite{Lovesey_multipoles_1, Lovesey_multipoles_2, Lovesey_book_right} for a complete derivation, while here we report the final result for the general $ll'$ matrix element: 
\begin{equation}
\begin{split}
\bigg \langle \frac{1}{2} \frac{j_0 (qr)}{q} \hat{\mathbf{q}} \wedge \boldsymbol{\nabla} \bigg \rangle_{ll'} &= \frac{1}{2} \left[ i I^2_{1, ll'} (q) \hat{\mathbf{q}} \wedge \langle \hat{\mathbf{r}} \rangle_{ll'} \right. \\ & \left. - I^{3}_{0,ll'} (q) \hat{\mathbf{q}} \wedge \langle \boldsymbol{\Omega}\rangle_{ll'} \right],    
\end{split}
\end{equation}
where the integrals 
\begin{equation}
    \begin{split}
    I^2_{L,ll'} (q) &= (2 L + 1) \int_0^{+\infty} \frac{j_L (k r)}{q^2 r} \left[ R_{n l} (r) \frac{d R_{n l'} (r)}{dr} \right. \\ & \left. - \frac{d R_{n l} (r)}{dr} R_{n l'} (r) \right] r^2 dr, 
    \end{split}
\end{equation}
\begin{equation}
    I^3_{L,ll'} (q) = \int_0^{+\infty} R_{nl} (r) \frac{j_L (qr)}{qr} R_{nl'} r^2 dr,
\end{equation}
have been introduced. 

\subsubsection{Magnetic scattering amplitude: summary}
\label{Mag_summary}
For the sake of clarity, here we summarize the main outcomes of the preceding sections. First, by means of a multipole expansion of both the spin and orbital contributions to $\mathbf{M} (\mk)$ in  Eq. \eqref{eq2}, we have demonstrated that local toroidal ($\mathbf{t}$) and quadrupole ($q$) ME multipoles can deflect neutrons. Second, we derived the spin and orbital ME form factors from the leading order terms of the expansion.

The leading-order terms of the multipole expansion of the single-atom contribution to $\mathbf{M} (\mk)$ can be written in the following compact formula:
\begin{widetext}
\begin{equation}
\label{eq14}
\begin{split}
\bigg \langle e^{i \mk \cdot \mathbf{r}_j} \, \left( \mathbf{s}_j - \frac{\hat{\mk} \wedge \boldsymbol{\nabla}_j}{q} \right) \bigg \rangle & \approx \sum_{l, l'} \left[ \underbrace{I^1_{0, ll'} (q) \langle \mathbf{s} \rangle_{ll'} + \left( I^1_{0, ll'} (q) + I^1_{2, ll'} (q) \right) \langle \mathbf{L} \rangle_{ll'}}_{\text{spin + orbital dipole approx.}} + \underbrace{i I^{1}_{1, ll'} (q) \left( \langle \widetilde{a} \rangle_{ll'} \hat{\mathbf{q}} + \frac{3}{2} \langle \widetilde{\mathbf{t}} \rangle_{ll'} \wedge \hat{\mathbf{q}} + 3 \langle \widetilde{Q} \rangle_{ll'} \hat{\mathbf{q}} \right)}_{\text{first-order ME multipole approx.}} \right. \\ & \left. + \frac{1}{2} i I^2_{1, ll'} (q)  \hat{\mathbf{q}} \wedge \langle \hat{\mathbf{r}} \rangle_{ll'} - \frac{1}{2} I^{3}_{0, ll'} (q) \hat{\mathbf{q}} \wedge \langle \mathbf{\Omega} \rangle_{ll'} \right].
\end{split}
\end{equation}
\end{widetext}
Thus far, in conventional magnetic neutron scattering, only the terms within the spin + orbital dipole approximation are considered and the inversion-breaking, higher-order terms neglected. However, if the ion resides on a Wyckoff position which breaks both space-inversion and time-reversal symmetries, ME multipoles can occur. To account for the contribution of these multipoles to the neutron scattering amplitude, we also have to include the higher-order terms pertaining to the ME multipole approximation. 

As we will discuss in more detail in Section~\ref{Experiment}, if these ME multipoles display cooperative long-ranged order, the scattered neutrons can collect into Bragg peaks. The intensity and reciprocal space location of the reflections will then depend on (i) the \textit{structure factor}, determined by the relative arrangement and orientation of the multipoles (i.e. the ME propagation and basis vectors), which can be predicted by a group-theory based symmetry analysis once the magnetic dipolar order is known, and (ii) the ME \textit{form factor}, to compute which, both the size of the multipoles and the radial integrals must be known (see Eq. \eqref{eq14} above). 

Interestingly, the magnetic dipole contributions are characterized by a form factor isotropic in reciprocal space, whereas the form factor of the ME multipolar contributions depends on the direction of the scattering wave vector, $\hat{\mathbf{q}}$. Furthermore, we remark that the dipole operators, $\mathbf{s}$ and $\mathbf{L}$, are parity-even, hence they allow only matrix elements with $l + l'$ even, while the ME multipole operators break inversion symmetry, therefore they have non-vanishing matrix elements for odd $l + l'$ only. 

\subsection{\texorpdfstring{Contribution of ME multipoles to $\mathbf{M} (\mathbf{q})$: \textit{ab initio} perspective}{me comp}}
\label{multipole_calc}

Having demonstrated theoretically how ME multipoles interact with neutrons, we now discuss how we can calculate, with first-principles computational techniques, the size and orientation of these multipoles, which can then be used to compute $\mathbf{M} (\mathbf{q})$ as discussed above. In the present work, we compute the matrix elements of the multipole tensor operators appearing in Eq. \eqref{eq7} by exploiting a decomposition of the density matrix computed with the full potential linearized augmented plane wave (FP-LAPW) density functional theory (DFT) method, as implemented in the \texttt{ELK} package \cite{multipole_decomposition, Elk_code}. 

The spin-spatial and spin-orbital multipole operators, $\mathbf{w}^{k p r}$ (here we adopt Racah's notation, to be consistent with Ref. \cite{multipole_decomposition}) are defined by a coupling of the spatial and spin indices, $k$ and $p$, of the double tensor $\mathbf{w}^{kp}$ (see Eqs. (23) and (26) of Ref. \cite{multipole_decomposition}), built as a combination of a rank-$k$ spatial tensor and a rank-$p$ spin tensor. $\mathbf{w}^{k p r}$ represents the most general case of the tensor operators $\mathbf{T}^{(\text{s})}_{L'}$ and $\mathbf{T}^{(\text{orb.})}_{L'}$ introduced in Eqs. \eqref{eq4} and \eqref{eq12}. In particular, the rank $r$ of the tensor $\textbf{w}^{kpr}$ corresponds to the rank $L'$ of $\mathbf{T}_{L'}$, the rank $k$ of the spatial part corresponds to the rank $L$ of $\mathbf{Y}_L$ in the definition of $\mathbf{T}_{L'}$, and the rank $p$ of the spin part corresponds to the rank of the tensor $\mathbf{s}$: in our case, $p = 1$ in Eq. \eqref{eq4} and $p = 0$ in Eq. \eqref{eq12}. As an example, the equivalent Racah's notation for the magnetic and ME tensor operators appearing in the leading-order terms of the multipole expansion of $\mathbf{M} (\mk)$, are reported in Table \ref{t1}. Here, the ME multipole tensors $\widetilde{a}$, $\widetilde{\mathbf{t}}$, and $\widetilde{Q}$ correspond to the irreducible spherical tensors $\textbf{w}^{110}$, $\textbf{w}^{111}$ and $\textbf{w}^{112}$, respectively.

The $ll'$ matrix element of the spherical component $t$ ($-r \le t \le r$) of $\mathbf{w}^{kpr}$ then reads: 
\begin{widetext}
\begin{equation}
\label{eq_multipoles}
\begin{split}
&\Big \langle w^{kpr}_{t} \Big \rangle_{l l'} = \sum_{m_l, m_{l'}} \sum_{\sigma_1, \sigma_2} \bigg \langle l, m_l, \frac{1}{2}, \sigma_1 \bigg \lvert w^{kpr}_t \bigg \lvert l', m_{l'}, \frac{1}{2}, \sigma_2 \bigg \rangle = \\ & \frac{1}{n_{kpr}} \sum_{m_l, m_{l'}} \sum_{\sigma_1, \sigma_2} \sum_{x, y} (-1)^{k-x} (-1)^{p-y} (-1)^{l-m_l} (-1)^{\frac{1}{2}-\sigma_1} \frac{1}{n_{ll'k}} \frac{1}{n_{\frac{1}{2}p}} \begin{pmatrix} k & r & p \\ -x & t & -y \end{pmatrix} \begin{pmatrix} l & k & l' \\ -m_l & x & m_{l'} \end{pmatrix} \begin{pmatrix} \frac{1}{2} & p & \frac{1}{2} \\ -\sigma_1 & y & \sigma_2 \end{pmatrix} \rho^{\nu \, \sigma_1 \sigma_2}_{l m_l l' m_{l'}}. 
\end{split}
\end{equation}
\end{widetext}
Here $n_{kpr}$ is a normalization factor coming from the coupling of the indices of the double tensor $\mathbf{w}^{kp}$ and is defined in Eq. (27) of Ref. \cite{multipole_decomposition}, whereas 
\begin{equation}
    n_{\frac{1}{2} p} = \frac{1}{\sqrt{(p+2)!}}
\end{equation}
corresponds to the factor $n_{sp}$ of Ref. \cite{multipole_decomposition} for the case $s=1/2$, and 
\begin{equation}
    n_{ll'k} = \frac{(l+l')!}{\sqrt{(l+l'-k)! \, (l+l'+k+1)!}}
\end{equation}
corresponds to the factor $n_{lk}$ of Ref. \cite{multipole_decomposition} generalized to the case $l \ne l'$. For a more detailed discussion about these normalization factors, we refer the interested reader to Refs. \cite{{van_der_laan_normalization, van_der_laan_3, ang_mom_theory}}.
The quantities in round brackets 
\begin{align}
&\begin{pmatrix} k & r & p \\ -x & t & -y \end{pmatrix}, \\ &\begin{pmatrix} l & k & l' \\ -m_l & x & m_{l'} \end{pmatrix}, \\
\text{and} &\begin{pmatrix} \frac{1}{2} & p & \frac{1}{2} \\ -\sigma_1 & y & \sigma_2 \end{pmatrix}
\end{align}
are Wigner-3j coefficients. Finally, $\rho^{\nu}$ contains the time-reversal even ($\nu=0$) and the time-reversal odd ($\nu=1$) parts of the density matrix and is defined from the full density matrix $\rho$ as: 
\begin{equation}
    \rho^{\nu} = \frac{1}{2} \left[ \rho + (-1)^{\nu} \mathcal{T} \rho \mathcal{T}^{\dagger} \right], 
\end{equation}
with $\mathcal{T}$ being the time-reversal operator. 

\subsection{\texorpdfstring{Contribution of ME multipoles to $\mathbf{M}(\mathbf{q})$: an experimental perspective}{me exp}}
\label{Experiment}

\begin{figure}[h!]
\includegraphics[width=0.49\textwidth]{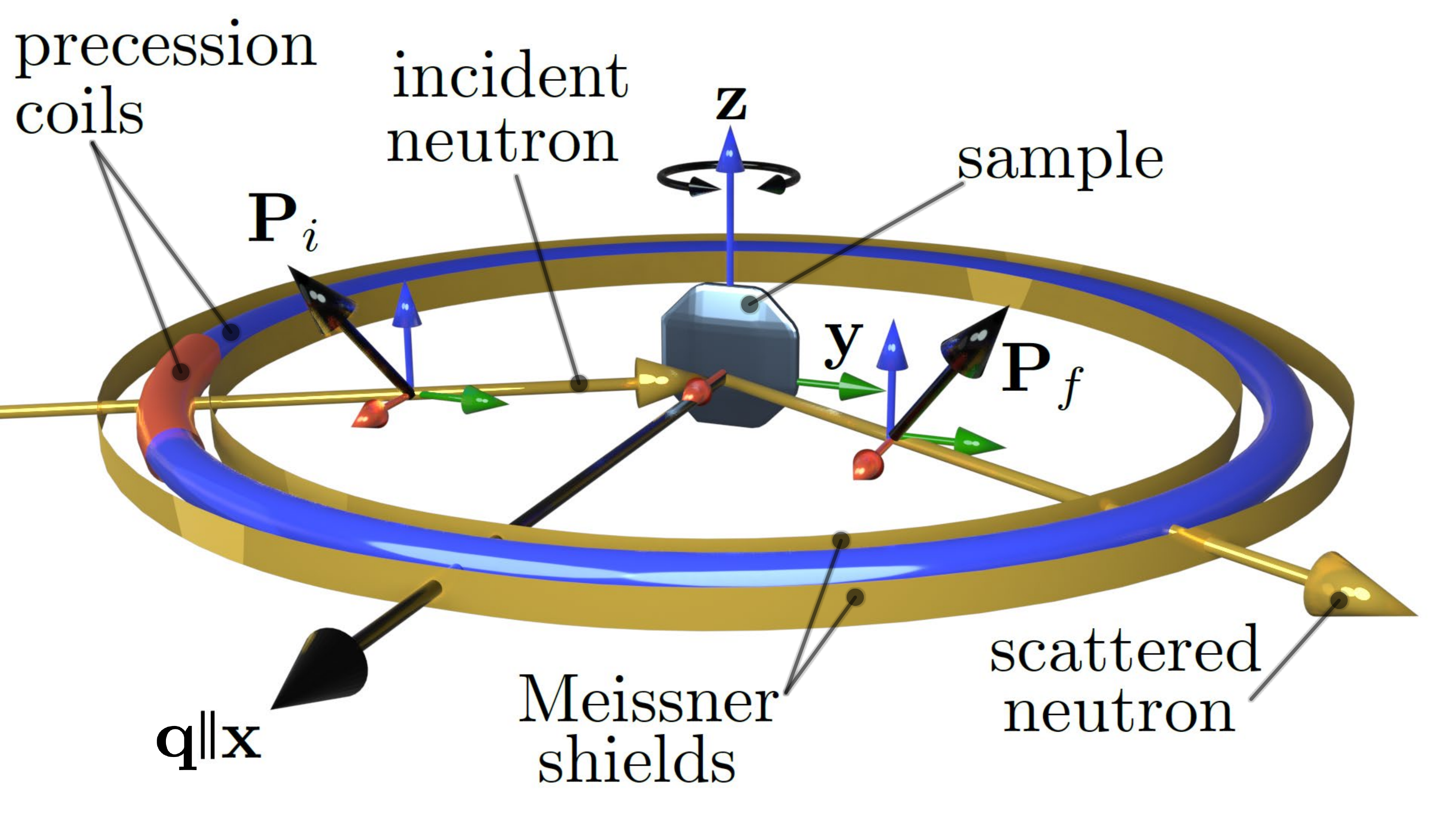}
\caption{The spherical neutron polarimetry set up as implemented on the D3 (ILL) beamline. Here, \textbf{x} is defined as the direction along the scattering vector \textbf{q}, \textbf{z} is normal to the horizontal scattering plane, and \textbf{y} completes the right-handed cartesian set. The sample is placed within a zero-field environment, afforded by two concentric cryogenically-cooled superconducting Meissner shields. Different  $(-h/2,k,h/2)$ reflections within the scattering plane can be accessed by rotating the sample about \textbf{z} and changing the size of the scattering vector \textbf{q}, which is set by the relative angle between the incident and scattered neutron wavevectors. The polarization of the incident neutron ($\textbf{P}_i$) can be controlled by a combination of nutator coil (not shown) and the incoming precession coil (in red). The outgoing precession coil (in blue) along with another nutator coil is used to analzye the polarization of the scattered neutrons ($\textbf{P}_f$).}
\label{fig:SNP}
\end{figure}

Thus far we have considered the contributions of local ME multipoles $\widetilde{a}$, $\widetilde{\mathbf{t}}$, and $\widetilde{q}$ to the magnetic interaction vector, $\mathbf{M} (\mathbf{q})$, theoretically and showed how to obtain them using DFT. Eqs. \eqref{eq_cross_sec}-\eqref{eq2} allow us to convert these calculated $\mathbf{M} (\mathbf{q})$ components into the neutron scattering intensity, which is the core physical quantity that is measured experimentally. The peaks found at specific positions in reciprocal space arise from the coherent scattering of neutrons from the long-ranged order of nuclei, magnetic dipoles and inversion-breaking ME multipoles. 
Here we consider purely magnetic and ME peaks, i.e. reflections that lack any nuclear and mixed nuclear-magnetic contribution. The magnetic dipolar and ME multipolar orders are defined by their propagation vectors, which we identify as $\textbf{q}_{\text{m}}$ and $\textbf{q}_{\text{ME}}$ respectively. As a consequence, the magnetic and ME reflections will be located at $\mathbf{q} = \mathbf{G}_{hkl} \pm \textbf{q}_{\text{m}}$ and $\mathbf{q} = \mathbf{G}_{hkl} \pm \textbf{q}_{\text{ME}}$, respectively, where $\mathbf{G}_{hkl}$ is the reciprocal lattice vector with Miller indices $(h,k,l)$. 

If $\textbf{q}_{\text{ME}} \ne \textbf{q}_{\text{m}}$, we have two well-distinguished families of peaks, one purely magnetic and one purely ME, in addition to the well-established nuclear structural peaks. 
In such a case, proving the existence of long-ranged order of ME multipoles is simply a matter of measuring the intensity (Eqs. \eqref{eq_cross_sec}-\eqref{eq_int}) of the ME peaks at the scattering vectors $\mathbf{q} = \mathbf{G}_{(h,k,l)} \pm \textbf{q}_{\text{ME}}$.
On the other hand, if $\textbf{q}_{\text{ME}} = \textbf{q}_{\text{m}}$, the two aforementioned families are located at the same positions in reciprocal space: in this case, the form factor contains contributions from both the magnetic dipoles and the ME multipoles. Since usually the ME multipoles are small compared to the main dipolar component in ME materials, it is difficult to single out the ME multipolar contribution to the magnetic scattering amplitude, $|\mathbf{M}_{\perp} (\mathbf{q})|$, from a conventional intensity experimental measurement with unpolarized neutron diffraction. However, due to the sensitivity to the direction of $\mathbf{M}_{\perp} (\mathbf{q})$, the analysis of the spin polarization of the scattered neutron beam yields more insights about the contribution of the ME multipolar order. Experimental measurements of the spin polarization of neutrons are typically carried out in a spherical neutron polarimeter, depicted in detail in Fig. \ref{fig:SNP}.

It is worthwhile noting that elastic scattering processes that only involve the nuclear structural peaks do not alter the orientation of the neutron spin polarization. Hence, in this case, the polarization of the scattered neutron, $\textbf{P}_f$, is the same as that of the incident neutron, $\textbf{P}_i$. On the other hand, scattering processes that involve magnetic interactions can change the orientation of the spin polarization of the neutron depending on its relative orientation to $\mathbf{M}_{\perp} (\mk)$. We analyze three different possibilities: 
\begin{enumerate}[label=(\roman*)]
    \item if $\textbf{P}_i \parallel \mathbf{M}_{\perp} (\mk)$ the spin orientation of the incident neutrons is preserved; \label{enum_a}
    \item if $\textbf{P}_i \perp \mathbf{M}_{\perp} (\mk)$ the spin polarization is flipped; \label{enum_b}
    \item if $\textbf{P}_i$ is neither parallel nor perpendicular to $\mathbf{M}_{\perp} (\mk)$, we can write 
        \begin{equation}
            \mathbf{P}_i = \mathbf{P}_{i \parallel} + \mathbf{P}_{i \perp},
        \end{equation}
        where $\mathbf{P}_{i \parallel}$ and $\mathbf{P}_{i \perp}$ are parallel and perpendicular to $\mathbf{M}_{\perp} (\mk)$, respectively. As a consequence, $\mathbf{P}_f$ reads 
\begin{equation}
    \mathbf{P}_f = \mathbf{P}_{i \parallel} - \mathbf{P}_{i \perp}.
\end{equation}
\label{enum_c}
\end{enumerate}
In both cases \ref{enum_a} and \ref{enum_b}, the scattered spin polarization lies in the same direction as the incident one, therefore $\lvert \mathbf{P}_f \cdot \mathbf{P}_i \lvert = 1$. On the other hand, in case \ref{enum_c}, $\mathbf{P}_f$ acquires a component perpendicular to $\mathbf{P}_i$; in turn, this results in a reduction of the component of $\mathbf{P}_f$ in the direction of $\mathbf{P}_i$, i.e. $\lvert \mathbf{P}_f \cdot \mathbf{P}_i \lvert < 1$.  

It is important to note, as discussed in more detail in Section \ref{exp_analysis} for the specific case of CuO, that a collinear magnetic order results in either $\mathbf{P}_i \parallel \mathbf{M}_{\perp} (\mk)$ or $\mathbf{P}_i \perp \mathbf{M}_{\perp} (\mk)$ in a typical spherical neutron polarimetry (SNP) setup, if the magnetic dipole moments point parallel or perpendicular to the scattering plane. In this case, the presence of additional non-collinear components and, possibly, hidden multipolar components, results in case \ref{enum_c} above and can therefore be detected by measuring a tilting of $\mathbf{P}_f$ compared to $\mathbf{P}_i$. Without loss of generality, $\textbf{P}_i$ is conventionally oriented along three principal directions, (\textbf{x}, \textbf{y}, \textbf{z}), where $\textbf{x}$ is along $\textbf{q}$, \textbf{z} is normal to the horizontal scattering plane, and the \textbf{y} axis is orthogonal to both \textbf{x} and \textbf{z} to complete the right-handed cartesian set (see Fig.~\ref{fig:SNP}). Similarly, $\textbf{P}_f$ can be reconstructed by analyzing the magnitude of the scattered neutron beam in the same three principal directions (\textbf{x}, \textbf{y}, \textbf{z}). 
	
The SNP results, for a given ($h,k,l$), can be compactly summarized as a polarization matrix $P$ with nine entries:
\begin{equation}
	P_{\alpha \beta} = \frac{n_{\alpha \beta} - n_{ \alpha \bar{\beta} }}{n_{ \alpha \beta } + n_{\alpha \bar{\beta}}},
\end{equation}
where $\alpha$ and $\beta$ denote the directions of the polarization of the incident and scattered neutron beams, respectively. $n_{\alpha \beta}$ and $n_{\alpha \bar{\beta}}$ are the number of scattered neutrons parallel and antiparallel to $\beta$ for incident neutron polarization along $\alpha$, respectively. For purely magnetic and ME reflections, the polarization matrix reads \cite{Boothroyd_book}
\begin{widetext}
\begin{equation}
\label{eq9}
    P = \begin{pmatrix} - \dfrac{(S^M_{yy} + S^M_{zz}) P_{i,x} + T^M_{yz}}{S^M_{yy} + S^M_{zz} + T^M_{yz} P_{i,x}} & 0 & 0 \\[0.5cm]
    - \dfrac{T^M_{yz}}{S^M_{yy} + S^M_{zz}} & \dfrac{S^M_{yy} - S^M_{zz}}{S^M_{yy} + S^M_{zz}} P_{i,y} & \dfrac{S^M_{yz}}{S^M_{yy} + S^M_{zz}} P_{i,y} \\[0.5cm] 
    - \dfrac{T^M_{yz}}{S^M_{yy} + S^M_{zz}} & \dfrac{S^M_{yz}}{S^M_{yy} + S^M_{zz}} P_{i,z} & - \dfrac{S^M_{yy} - S^M_{zz}}{S^M_{yy} + S^M_{zz}} P_{i,z} \end{pmatrix},
\end{equation}
\end{widetext}
where $S^M_{yy}$, $S^M_{zz}$, $S^M_{yz}$, and $T^M_{yz}$ are computed from the components of $\mathbf{M}_{\perp} (\mk)$ along the principal directions \textbf{x}, \textbf{y}, \textbf{z} in the following way: 
\begin{align}
    \label{eq15}
    S^M_{yy} & = |M_{\perp y} (\mk)|^2, \\ 
    S^M_{zz} & = |M_{\perp z} (\mk)|^2, 
\end{align}
\begin{align}
    S^M_{yz} & = M^*_{\perp y} (\mk) M_{\perp z} (\mk) + M_{\perp y} (\mk) M^*_{\perp z} (\mk), \\ 
    \label{eq16}
    T^M_{yz} & = M^*_{\perp y} (\mk) M_{\perp z} (\mk) - M_{\perp y} (\mk) M^*_{\perp z} (\mk),
\end{align}
and $P_{i,x}$, $P_{i,y}$, and $P_{i,z}$ are the polarization values for an incident beam polarized along $\mathbf{x}$, $\mathbf{y}$, or $\mathbf{z}$, respectively.

\subsection{Summary}

As a summary, we review our complete workflow, starting from the DFT-calculated ME multipoles to the computation of the neutron scattering cross sections. The predictions from this workflow will (i) aid in the identification of the reciprocal-space positions of ME peaks for a decisive proof of long-ranged order of ME multipoles and (ii) support the analysis and interpretation of experimental measurements.

\begin{enumerate}
\item Calculate the ground state crystal structure and magnetic order of the material using DFT. Benchmark against any previous experimental reports. 

\item Compute the size of the multipoles $\widetilde{a}$, $\widetilde{\mathbf{t}}$, and $\widetilde{Q}$ following Eq. \eqref{eq_multipoles}, as implemented in the \texttt{ELK} DFT code. Note that the ME multipole order does not need to be specified as an input, as it follows by symmetry from the dipolar order (see the discussion for CuO in Section \ref{theo_analysis}).

\item Insert $\widetilde{a}$, $\widetilde{\mathbf{t}}$, and $\widetilde{Q}$ into Eq. \eqref{eq14} above to obtain the single-ion contribution to $\mathbf{M} (\mathbf{q})$. 

\item Use Eq. \eqref{eq2} to compute the $\mathbf{M} (\mathbf{q})$ of the magnetic unit cell. Note that when computing the sum over the ions, the correct relative arrangement of the multipoles must be respected. 

\item Compute $\mathbf{M}_{\perp} (\mathbf{q})$ (Eq. \eqref{eq1}). 

\item From $\mathbf{M}_{\perp} (\mathbf{q})$, compute either $I(\mathbf{q}) \propto \lvert \mathbf{M}_{\perp}(\mathbf{q}) \lvert^2$ or $S^M_{yy}$, $S^M_{zz}$, $S^M_{yz}$, and $T^M_{yz}$ (Eqs. \eqref{eq15}-\eqref{eq16}), based on the discussion in Section \ref{Experiment}. 

\item Compare with experimental results available at specific scattering wave vectors $\mathbf{q}$.

\end{enumerate}

\section{\texorpdfstring{A specific case: C\lowercase{u}O}{CuO}}
\label{CuO}

In the preceding sections, we have discussed how ME multipoles contribute to the neutron scattering cross section from theoretical, computational and experimental perspectives. We next illustrate this contribution with a specific example.  

We choose CuO as it fulfils the two necessary conditions to host ME multipoles, namely (i) that the ions (Cu and O) reside on Wyckoff positions that lack inversion symmetry and (ii) time-reversal symmetry is broken below $T$ = 229.3\,K. CuO displays three magnetic phases, AF1, AF2, and AF3 \cite{Qureshi_2020_CuO_SNP}, in zero magnetic field: in this work we study the first, which occurs below 213\,K. 
An earlier study of AF1 CuO with resonant x-ray scattering (REXS) provided some indication for orbital currents interpreted as toroidal multipoles \cite{me_multipoles, toroidal_CuO}, although the observed REXS signal has also been attributed to birefringence effects \cite{PhysRevB.86.220101}. To reach a resolution, further experimental searches for these ME multipoles are required. Moreover, although the CuO ME multipoles have been modeled with a model wave function~\cite{Lovesey_multipoles_1}, an \textit{ab initio} calculation of the multipoles, which is crucial for a direct comparison to experimental results, is still lacking.

To address these open questions we: (i) calculate the size and orientation of the various ME multipoles in CuO with DFT; (ii) convert these multipoles to neutron scattering cross sections; (iii) search for the existence of ME multipoles by measuring these neutron scattering cross sections with high precision neutron diffraction experiments.

\begin{figure*}[t!] 
\includegraphics[width=0.9\textwidth]{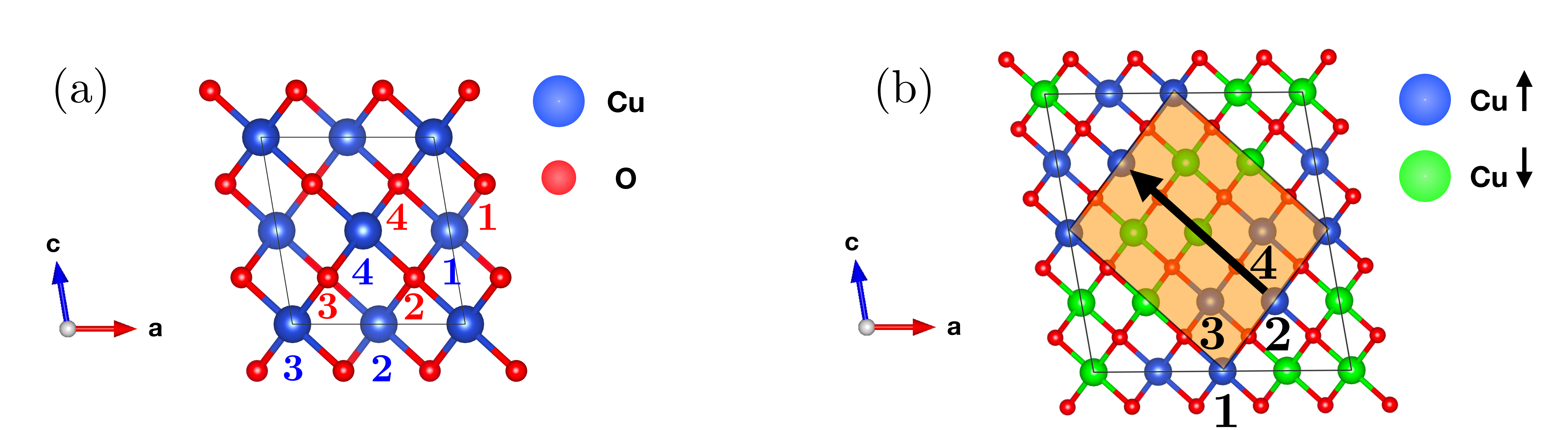}
\caption{\label{fig:unit_cells} CuO crystallographic and magnetic unit cells. (a) Crystallographic unit cell: view of the $ac$ plane. Cu (O) atoms, occupying the Wyckoff site $4e$ of the $C 2/c$ space group are numbered in blue (red). (b) Magnetic conventional cell of the AF1 phase, with propagation vector $\textbf{q}_{\text{m}} = (-1/2, 0, 1/2)$ identified by the black arrow. The primitive magnetic cell, described by the magnetic space group $P 2_1 / c$, is highlighted by the orange shaded area. Cu ions, occupying the Wyckoff site $4e$, are numbered in black.}
\end{figure*}

\subsection{Methods}

\subsubsection{Computational details} 
\label{theo_methods}

\textit{Ab initio} calculations, including structural relaxations, were performed in the local spin density approximation (LSDA) as implemented in the Quantum \texttt{ESPRESSO} (QE) \cite{QE_1, QE_2} and \texttt{thermo\_pw} \cite{thermo_pw} packages. The ions were described by scalar relativistic ultrasoft \cite{US_Vanderbilt} pseudopotentials (PP) with 4$s$ and 3$d$ valence electrons for Cu (PP \texttt{Cu.pz-dn-rrkjus\_psl.1.0.0.UPF} from pslibrary 1.0.0 \cite{pslibrary, pslibrary_2}) and with 2$s$ and 2$p$ valence electrons for O (PP \texttt{O.pz-n-rrkjus\_psl.1.0.0.UPF} from pslibrary 1.0.0). To appropriately describe the antiferromagnetic insulating ground state we introduced a Hubbard $U$ correction, following the simplified scheme by Cococcioni and de Gironcoli \cite{U_cococcioni}, on Cu $d$ states, with $U = 9$ eV and $J = 2.5$ eV, a choice motivated by previous works on similar compounds \cite{CuO_cococcioni} and by the agreement with experiments for the electronic band gap, the lattice constants, and the Cu magnetic moments. The equilibrium structure was modeled with the monoclinic primitive cell, described by the space group $C2/c$, shown in Fig. \ref{fig:unit_cells}(a). Our calculated LSDA+$U$ equilibrium lattice constants are $a = 4.59$ \AA, $b = 3.35$ \AA, $c = 5.04 $ \AA, and $\beta = 99.4^{\circ}$, $2 \%$ ($a$, $b$, and $c$) and $0.15 \%$ ($\beta$) smaller than the reported experimental values \cite{CuO_structure}. The magnetic ordering is antiferromagnetic, with propagation vector $\textbf{q}_{\text{m}} = (-1/2, 0, 1/2)$; as a consequence, the size of the magnetic cell is four times the crystallographic primitive cell (see Fig. \ref{fig:unit_cells}(b)). The smallest possible magnetic cell, identified by the orange shaded area in Fig. \ref{fig:unit_cells}(b), is described by the magnetic space group $P 2_1/c$. The pseudowave functions (charge density) were expanded in a plane-wave basis set with a kinetic energy cut-off of 80 (320) Ry. The Brillouin Zone (BZ) was sampled by means of a $\Gamma$-centered uniform Monkhorst-Pack mesh \cite{Monkhorst_Pack_mesh} with $4 \times 8 \times 4$ $\mathbf{k}$ points.  

The ME multipoles were computed by decomposing the density matrix into spherical tensor moments, as described in Section \ref{multipole_calc}, as implemented in a customized version of the FP-LAPW code \texttt{ELK} \cite{Elk_code}, based on version 3.3.17. Self-consistent ground-state calculations were performed at the QE-calculated LSDA + $U$ lattice parameters mentioned earlier. The APW wave functions were expanded in a spherical harmonic basis set, with a cut-off $l_{\text{max (apw)}} = 8$, and the BZ was sampled with a $4 \times 8 \times 4$ uniform, $\Gamma$-centered Monkhorst-Pack mesh.   

Finally, the radial integrals appearing in Eq. \eqref{eq14} were evaluated using the model radial wave functions computed with the Roothaan-Hartree-Fock method and reported in Ref. \cite{radial_functions}.

\subsubsection{Experimental details} 
\label{exp_methods}

Single crystals of CuO were grown via the high-temperature solution growth method as described in Refs.~\cite{Qureshi_2020_CuO_SNP,wang_magnetoelectric_2016}. The neutron diffraction experiments on CuO were performed on the D3 diffractometer~\cite{LELIEVREBERNA2005141} at the Institut Laue-Langevin (ILL) and the TASP triple-axis spectrometer~\cite{Boni_1996_TASP} at the Swiss Spallation Neutron Source (SINQ). To assess the robustness of our results, we used two different implementations of SNP, namely the CryoPAD \cite{Tasset_1989_SNP,Lelievreberna_2005_SNP} and MuPAD \cite{Janoschek_2007_SNP} polarimeters. 

On the D3 diffractometer, the incident neutron wavelength ($\lambda = 0.832$\AA) was selected with a ferromagnetic Cu$_2$MnAl (111) Heusler monochromator, which also polarizes the neutrons along the incident wavevector. The incident and scattered neutron polarization were controlled with a combination of nutator and precession coils. The sample was encased within two cryogenically-cooled Meissner shields to minimise the neutron depolarization from the external magnetic field. The scattered neutrons were filtered with a field-polarized $^3$He spin filter cell (that is changed every $\sim$24 hours) and measured with a $^3$He detector. 
On the TASP spectrometer, the wavelength of incident neutrons ($\lambda = 3.14$ \AA) was chosen with a pyrolytic graphite (002) monochromator. The incident neutrons were polarized with a polarizing bender. The polarization of the incident and scattered beam were manipulated with a set of four nutator coils. The sample was placed within a mu-metal alloy shield to reduce the depolarization of the neutrons. To maximize the scattered neutron intensity, the experiment was performed with a two-axis spectrometer mode, where the scattered neutron beam was not analyzed. 

Measurements on both beamlines were performed at $T = 1.5$\,K where CuO is well within the AF1 phase. 
Different crystals from the same batch were used for the two experiments to test for reproducibility. To reduce the systematic uncertainties that are inherent in neutron polarimeters, the measurements were averaged over the Friedel pairs, the equivalent reflections and the two polarities of the incident neutron beam along each principal direction. 

Prior reports on SNP measurements of CuO \cite{Brown_1991_CuO_SNP,Ain_1992_CuO_SNP,Babkevich_2012_CuO_SNP,Qureshi_2020_CuO_SNP} were performed with the crystal \textbf{b} axis oriented perpendicular to the horizontal scattering plane (\textbf{q}$\perp$\textbf{b}), where reflections with mixed magnetic and ME multipoles were not accessible (see a more detailed discussion in Section \ref{exp_analysis}). To be sensitive to the contribution of the long-range ME multipolar order to the scattered neutron intensity, we need to choose a scattering vector \textbf{q}=$(h,k,l)$ with $k$$\neq$0. Hence we aligned the CuO single crystal with the $(0,1,0)$ and $(-1/2, 0, 1/2)$ reciprocal vectors within the horizontal scattering plane so that \textbf{q}=$(-h/2, k, h/2)$ reflections were accessible in the two-circle diffractometer set-up. The single crystals were aligned with the neutron (OrientExpress~\cite{Ouladdiaf_2006_OrientExpress}) and x-ray (Multiwire) Laue diffractometers.

\subsection{Results and discussion} 

\subsubsection{DFT results and symmetry analysis}
\label{theo_analysis}

As remarked in Section \ref{theory}, the main ingredients that affect the ME structure factor and form factor are the multipolar order, dictated by the magnetic point symmetry, the size of the multipoles, and the radial integrals. \textit{Ab initio} DFT techniques are ideal for assessing the first two aspects, whereas the radial integrals must be evaluated with model wave functions, as described in Section \ref{theo_methods}.

\begingroup
\begin{table}[t]
\begin{tabular}{|c|c|c|c|c|c|c|} 
\hline 
$2/m$ & \multirow{2}{*}{$m_x, m_z$} & \multirow{2}{*}{$m_y$} & \multirow{2}{*}{$\widetilde{t}_x, \widetilde{t}_z$} & \multirow{2}{*}{$\widetilde{t}_y$} & $\widetilde{q}_{x^2-y^2},$ & \multirow{2}{*}{$\widetilde{q}_{xy}, \widetilde{q}_{yz}$} \\
Irrep & & & & & $\widetilde{q}_{z^2}, \widetilde{q}_{xz}$ & \\
\hline 
$A_g$ & $+ - - +$ & $+ + + +$ & $+ + - -$ & $+ - + -$ & $+ - + -$ & $+ + - -$ \\ 
\hline
$A_u$ & $+ + - -$ & $+ - + -$ & $+ - - +$ & $+ + + +$ & $+ + + +$ & $+ - - +$ \\ 
\hline
$B_g$ & $+ + + +$ & $+ - - +$ & $+ - + -$ & $+ + - -$ & $+ + - -$ & $+ - + -$ \\ 
\hline 
$B_u$ & $+ - + -$ & $+ + - -$ & $+ + + +$ & $+ - - +$ & $+ - - +$ & $+ + + +$ \\
\hline 
\end{tabular}
\caption{Allowed arrangements of magnetic dipoles, toroidal moments, and quadrupole moments on the $4e$ Wyckoff position for each irrep of the point group $2/m$. The magnetic order of CuO corresponds to the irrep $A_g$.}
\label{t2}
\end{table}
\endgroup

\begingroup
\begin{table}[t]
\begin{tabular}{|c|c|c|} 
\multicolumn{3}{c}{Magnetic dipoles ($\mu_{\text{B}}$)} \\
\hline 
Component & Cu & O \\
\hline
$m_x$ & 0.014 & 0.003 \\
\hline
$m_y$ & 0.675 & 0.102 \\  
\hline
$m_z$ & 0.016 & 0.003 \\
\hline 
\multicolumn{3}{c}{} \\
\multicolumn{3}{c}{Toroidal moments ($\times 10^{-2} \mu_{\text{B}}$)} \\
\hline
Component & Cu & O \\
\hline
$\widetilde{t}_x$ & 0.061 & 2.528 \\ 
\hline 
$\widetilde{t}_y$ & 0.001 & 0.049 \\ 
\hline
$\widetilde{t}_z$ & 0.006 & 0.251 \\ 
\hline 
\multicolumn{3}{c}{} \\
\multicolumn{3}{c}{ME quadrupoles ($\times 10^{-2} \mu_{\text{B}}$)} \\
\hline
Component & Cu & O \\
\hline
$\widetilde{q}_{xy}$ & \hspace{0.14cm} 0.005 & \hspace{0.14cm} 0.224 \\ 
\hline 
$\widetilde{q}_{yz}$ & \hspace{0.14cm} 0.080 & \hspace{0.14cm} 3.095 \\
\hline 
$\widetilde{q}_{z^2}$ & $-0.027$ & $-1.151$ \\
\hline 
$\widetilde{q}_{xz}$ & $-0.004$ & \hspace{0.14cm} 0.075 \\
\hline
$\widetilde{q}_{x^2 - y^2}$ & $-0.033$ & $-1.860$ \\ 
\hline
\end{tabular} 
\caption{Sizes of the components of magnetic dipoles, and of the angular part of toroidal moments and ME quadrupoles, at the Cu and O sites. The $x$ and $y$ axes correspond to the directions given by the $\mathbf{a}$ and $\mathbf{b}$ lattice vectors.}
\label{t3}
\end{table}
\endgroup

Interestingly, even though the net ME effect is symmetry-forbidden in the AF1 phase of CuO since the crystal is symmetric by inversion, local ME multipoles are allowed for both Cu and O ions, which reside on the $4e$ Wyckoff position in the $P 2_1/c$ setting, because the atomic sites lack inversion and time-reversal symmetry. 

Our DFT calculations show that the Cu and O ME multipoles arrange antiferroically, with a different pattern compared to the magnetic dipoles (see Table \ref{t2}). Nevertheless, the ME propagation vector, $\textbf{q}_{\text{ME}}=(-1/2,0,1/2)$, is identical to the magnetic propagation vector $\mathbf{q}_{\text{m}}$, hence the ME reflections appear at the same reciprocal space vectors as the magnetic reflections, given by $\textbf{q}=\mathbf{G}_{hkl} \pm \textbf{q}_\mathrm{m}$, where $h$, $k$, and $l$ correspond to the Miller indices of allowed structural Bragg peaks. The allowed reflections can be determined by calculating the structure factor from the relative arrangement and orientation of the ME multipoles across the magnetic cell. The allowed arrangements can be expressed elegantly in terms of the irreducible representations (irreps) of the crystallographic point group by analyzing how each multipole transforms under the allowed symmetry operations of the crystal (see e.g.\ \cite{me_multipoles}). 

Focusing on Cu ions, we find all possible symmetry-allowed magnetic and multipolar structures, of which only one corresponds to the magnetic ground state of CuO. Each symmetry-allowed arrangement follows a different active irrep among the four 1-dimensional irreps of $2/m$ ($A_g$, $A_u$, $B_g$, $B_u$), and is summarized in Table \ref{t2}. Here, the ($\pm\pm\pm\pm$) refer to the relative pattern of the Cu magnetic dipole moments and ME multipoles at sites $1$-$4$, defined with respect to the primitive magnetic unit cell as shown in Fig.~\ref{fig:unit_cells}(b). Earlier studies of CuO \cite{Brown_1991_CuO_SNP,Ain_1992_CuO_SNP,Babkevich_2012_CuO_SNP,Qureshi_2020_CuO_SNP} in the AF1 phase show that the dipolar magnetic configuration has the collinear antiferromagnetic order along $\mathbf{b}$ depicted in Fig. \ref{fig:unit_cells} (b), transforming as the $A_g$ irrep. Then, assuming that the multipolar order can be described by a single irrep, the ME toroidal moments within the $a$-$c$ plane and the $xy$ and $yz$ quadrupoles have a $(++--)$ arrangement, whereas the component of $\mathbf{t}$ along $\mathbf{b}$ and the $x^2-y^2$, $z^2$, and $xz$ quadrupoles follow a $(+-+-)$ order. Such arrangements are confirmed by our DFT calculations. 

The predicted sizes of the magnetic dipoles and the ME multipoles on the Cu and O atomic sites are reported in Table \ref{t3}. The main dipolar magnetic order is collinear, points along $\mathbf{b}$, and is carried by the Cu atoms. The O atoms contribute to the collinear order as well, with a magnetic dipole about 6 times smaller than Cu. Remarkably, both Cu and O also show small non-collinear components (i.e. $\perp \mathbf{b}$) which, to the best of our knowledge, have not been reported in previous theoretical or experimental works on the AF1 phase. Concerning the ME multipoles, the Cu atoms carry small toroidal moments and quadrupole moments as expected by symmetry, whereas the ME multipoles on the O atoms are much larger, in the same ballpark as the magnetic dipoles. It is noteworthy that the O atoms show the most prominent multipoles, even though their magnetic dipole is much smaller than Cu. 

\subsubsection{Experimental results and analysis}
\label{exp_analysis}

\begin{figure}[th]
\includegraphics[width=0.48\textwidth]{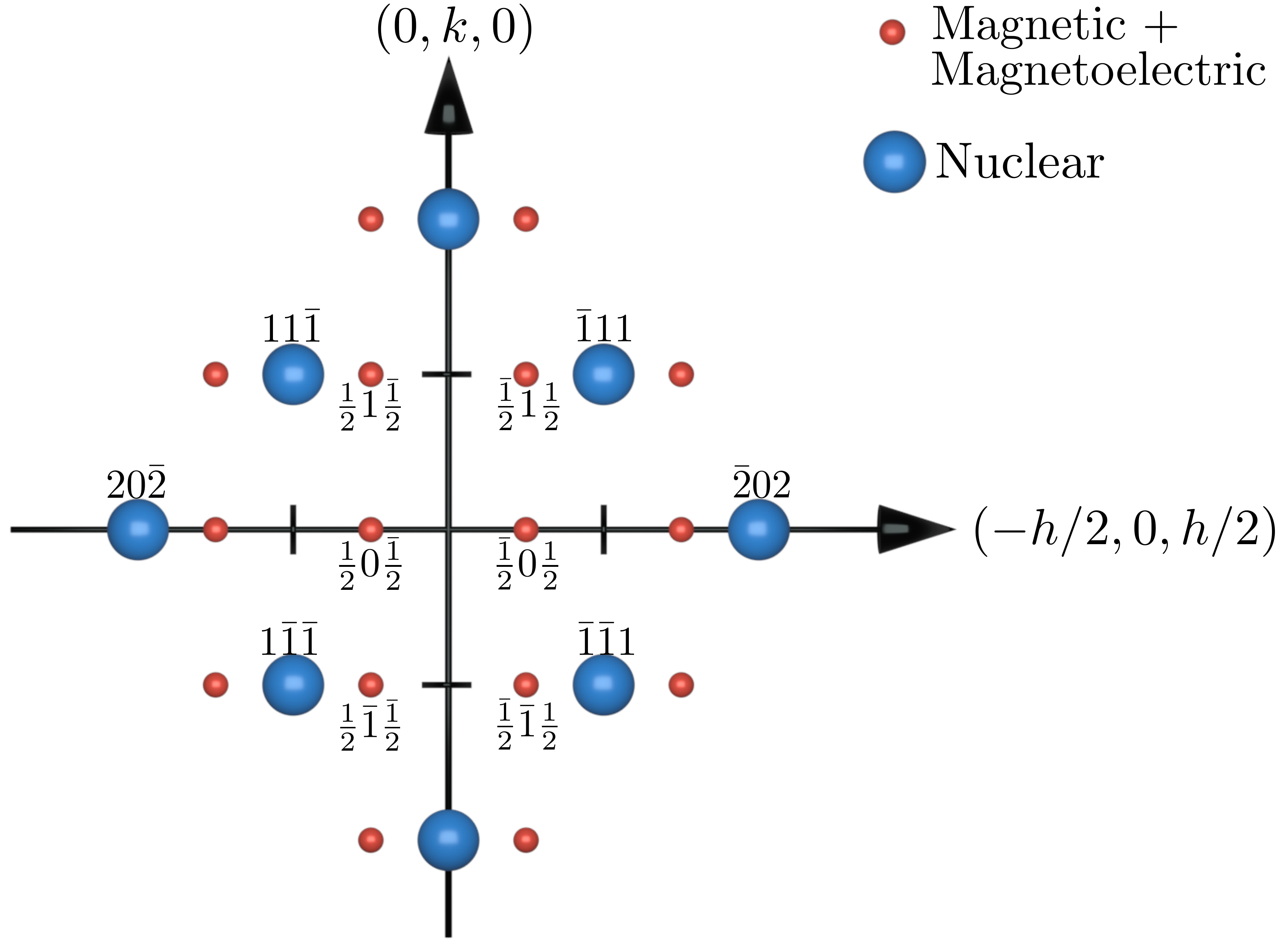}
\caption{\label{fig:Horizontal_Scatt_Plane} Allowed reflections in the horizontal scattering plane for CuO. The nuclear and mixed (magnetic + ME) reflections are shown in blue and red circles, respectively.} 
\end{figure}

As mentioned above, the magnetic structure of CuO is described by the magnetic propagation vector $\textbf{q}_{\text{m}} = (-1/2, 0, 1/2)$, and moreover the ME multipolar arrangement has a propagation vector $\textbf{q}_{\text{ME}} = \textbf{q}_{\text{m}}$. As a consequence, the allowed reflections can be divided, at first, into two families, as shown in Fig. \ref{fig:Horizontal_Scatt_Plane}: (i) structural, appearing at $\mathbf{q} = (h, k, l)$, and (ii) mixed magnetic and ME, appearing at $\mathbf{q} = (h - 1/2, k, l + 1/2)$. 

\begin{figure}[t]
\includegraphics[width=0.42\textwidth]{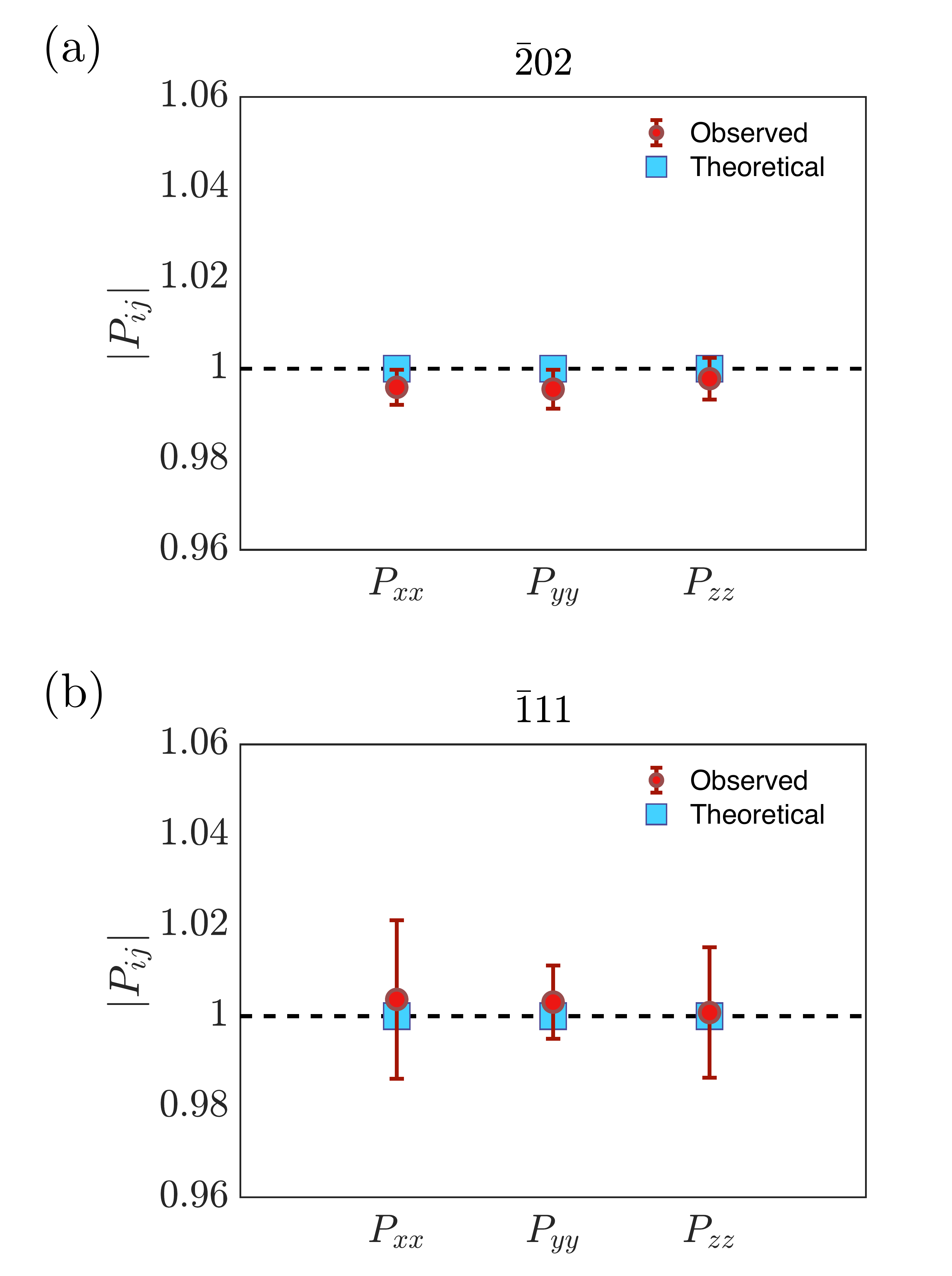}
\caption{Predicted and experimental values of the diagonal entries of the polarization matrix for the nuclear reflections ($\bar{2} 0 2$) and ($\bar{1} 1 1$).}
\label{fig_nuclear_peaks}
\end{figure}

To assess the presence of ME multipoles in CuO, we need to determine, experimentally and computationally, the magnitude and direction of the scattered neutron beam for a given incident neutron polarization. In the following, we present a polarization analysis for some selected allowed reflections. First, we take some purely nuclear reflections, indicated by the blue dots in Fig. \ref{fig:Horizontal_Scatt_Plane}, to assess the quality of the experimental setup; as stated above, they are characterized by $\mathbf{P}_f = \mathbf{P}_i$, therefore the polarization matrix is the identity matrix, $P = \text{diag} (1,1,1)$. The experimental results for the reflections $(\bar{2}, 0, 2)$ and $(\bar{1}, 1, 1)$ are reported in Fig. \ref{fig_nuclear_peaks} and show that the experimental setup provides accurate polarization measurements. Second, we consider two families of magnetic and ME reflections in the scattering plane $(-h/2, k, h/2)$, corresponding to the $k = 0$ and $k \ne 0$ cases, respectively. Below we take one representative from each family, and discuss the theoretical predictions and experimental results concerning the diagonal entries of $P$. 

\begin{figure}[t]
\includegraphics[width=0.45\textwidth]{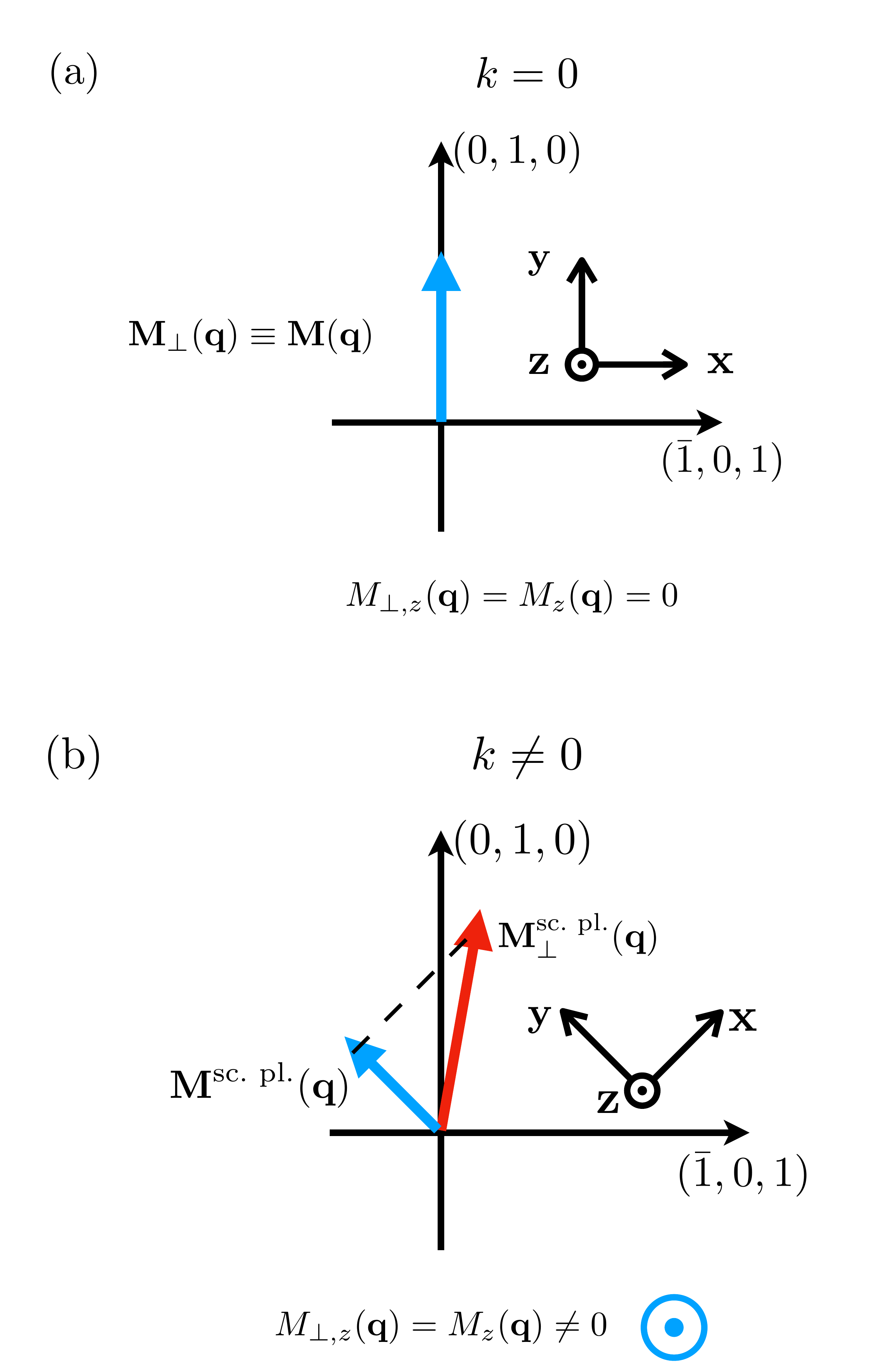}
\caption{Orientation of $\mathbf{M}_{\perp} (\mk)$ (red arrows) and $\mathbf{M} (\mk)$ (blue arrows) for the two cases discussed in the main text, $k = 0$ (panel (a)) and $k \ne 0$, specifically $\mathbf{q} = (-1/2, 1, 1/2)$ (panel (b)). Both the components parallel and perpendicular to the scattering plane (sc.\ pl.) are shown. The $\textbf{x}$, $\textbf{y}$, $\textbf{z}$ Blume-Maleev polarization axis system, showing the directions of the polarization of the incident neutron beam in a standard spherical neutron polarimetry setup, is shown in black for each case.}
\label{fig_m_perp}
\end{figure}

\begin{figure*}[ht]
\includegraphics[width=0.95\textwidth]{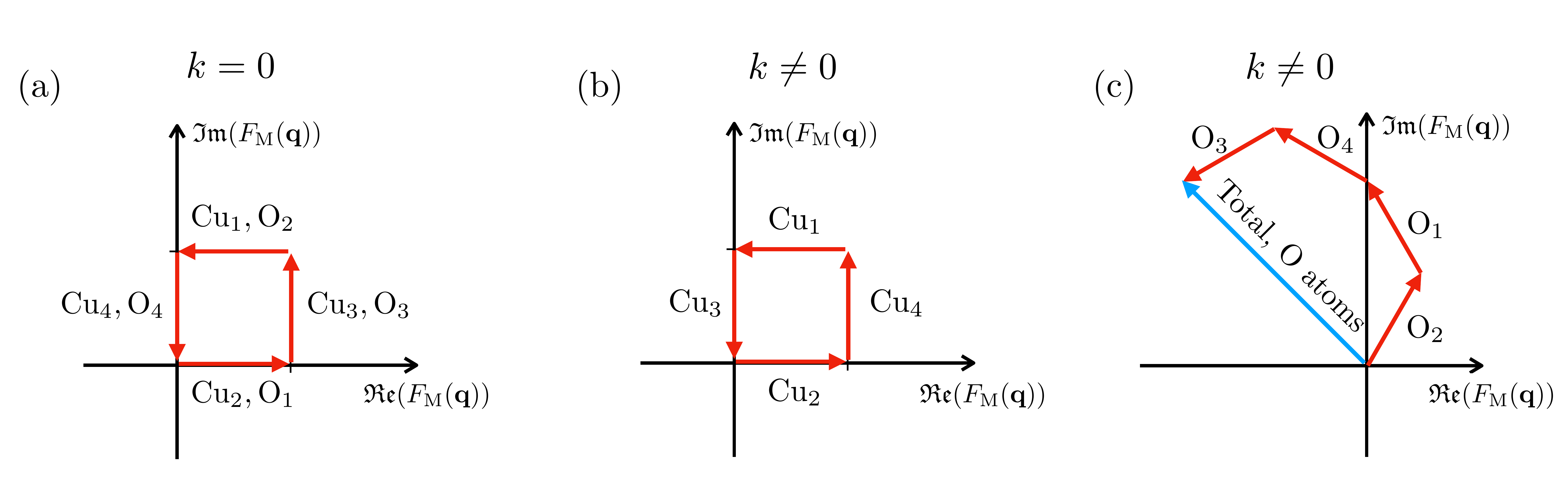}
\caption{Analysis of the structure factor of the non-collinear magnetic dipole components, computed for each Cu and O atom, numbered according to Fig. \ref{fig:unit_cells}(a). The magnetic structure factor is defined as $F_{\text{M}} (\mathbf{q}) = \sum_j (\pm) e^{i \mathbf{q} \cdot \mathbf{d}_j}$, where $\mathbf{d}_j$ identifies the position of the $j$-th atom in the unit cell, and the $(\pm)$ sign in front is the sign of the non-collinear dipole component of the atom $j$. (a) Cu and O structure factors in the $k = 0$ case, where $F_{\text{M}} (\mathbf{q}) = 0$. (b) Cu structure factors in the $k \ne 0$ case, where $F_{\text{M}} (\mathbf{q})$ is also 0. (c) O structure factors in the $k \ne 0$ case, where $F_{\text{M}} (\mathbf{q}) \ne 0$, denoted by the blue arrow.}
\label{fig_strf}
\end{figure*}

\subsubsection*{\texorpdfstring{Case I: $k = 0$}{k = 0}} 

\begin{figure}[t!]
\includegraphics[width=0.45\textwidth]{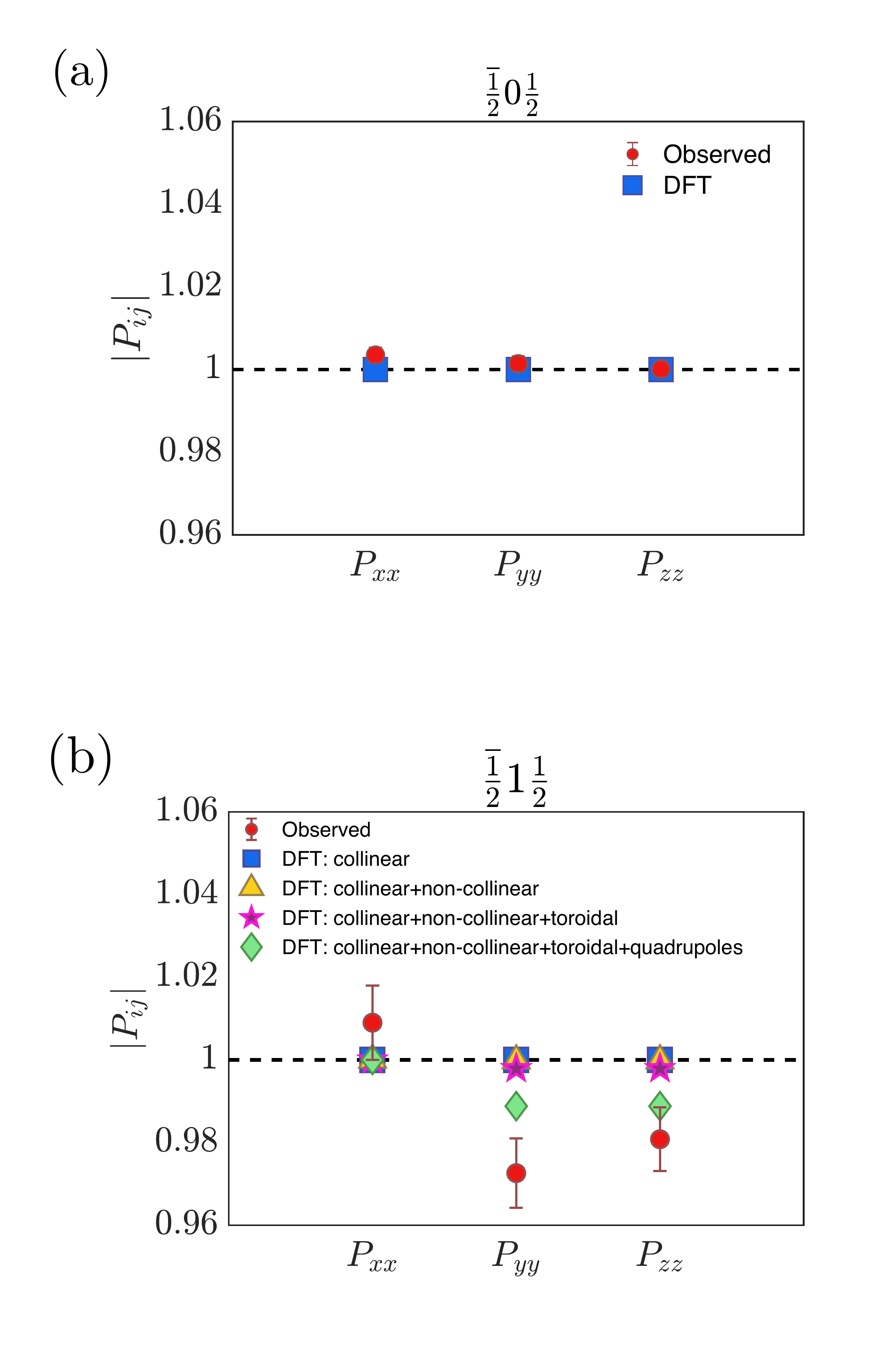}
\vspace*{2mm}
\caption{Magnitude of the diagonal entries of the polarization matrix. First-principles predictions and experimental data for (a) $\mathbf{q} = (-1/2, 0, 1/2)$. and (b) $\mathbf{q} = (-1/2, 1, 1/2)$. DFT results are shown starting from purely collinear dipolar magnetic order (blue squares), and by progressively adding the non-collinear components (yellow triangles), and the higher-order toroidal (magenta stars) and quadrupolar (green diamonds) contributions.}
\label{fig_Pyy_Pzz}
\end{figure}

According to our theoretical predictions and experimental measurements, $\mathbf{M}_{\perp} (\mathbf{q})$ always lies in the scattering plane and is always parallel or perpendicular to $\mathbf{P}_i$, as shown in Fig. \ref{fig_m_perp} (a). As a consequence, the polarization of the scattered neutron beam, $\mathbf{P}_f$, is always parallel or anti-parallel to $\mathbf{P}_i$. This feature can be explained by inspecting each source of magnetic scattering separately. First, the main collinear component of the Cu and O magnetic dipoles lies in the scattering plane, hence it will contribute only to $S^M_{yy}$ (see Fig. \ref{fig_m_perp} (a) for the relative orientation of the scattering plane to the Blume-Maleev principal axes). Second, the non-collinear spin components, which yield a component of $\mathbf{M} (\mk)$ perpendicular to the scattering plane (i.e. $M_z (\mk) \ne 0$, hence a non-zero $S^M_{zz}$ and, in turn, $|P_{yy}|, |P_{zz}| < |P_{xx}|$), are suppressed by a vanishing structure factor, illustrated in Fig. \ref{fig_strf} (a). Third, neither toroidal nor quadrupole moments do not contribute to $M_z (\mk)$ in this specific instance. In particular, since both $\mathbf{t}$ and $\mathbf{q}$ lie in the $a$-$c$ plane, the toroidal contribution to $\mathbf{M} (\mk)$, proportional to $\mathbf{t} \wedge \mathbf{q}$, is parallel to $\mathbf{b}$, and hence to the principal axis $y$. Concerning the ME quadrupole term, following Eq. \eqref{eq6} and Table \ref{t3} only $q_{xz}$ and $q_{z^2}$ would contribute to a non-vanishing $M_z (\mk)$, however both components are suppressed by a vanishing structure factor. In Fig. \ref{fig_Pyy_Pzz} (a) we report the experimental results and the DFT predictions obtained at $\mathbf{q} = (-1/2, 0, 1/2)$. Our measurements confirm the predicted $\lvert \mathbf{P}_f \cdot \mathbf{P}_i \lvert = 1$, with $|P_{xx}| = |P_{yy}| = |P_{zz}| = 1$ within the error bars. 

\subsubsection*{\texorpdfstring{Case II: $k \ne 0$}{non-zero k}} 

In this case, our DFT calculations predict a non-zero out-of-plane component of $\mathbf{M}_{\perp} (\mk)$, which results in $S^{M}_{zz} \ne 0$ in Eq. \eqref{eq9} and hence $|\mathbf{P}_f \cdot \mathbf{P}_i| < 1$, or equivalently $|P_{yy}|, |P_{zz}| < |P_{xx}|$. This component arises from the following contributions: 
\begin{enumerate}[label=(\roman*)]
    \item non-collinear magnetic dipolar components. Only the O atoms provide a non-vanishing contribution, since the structure factor for the components of the magnetic dipoles along the $\mathbf{a}$ and $\mathbf{c}$ crystal axes is non-zero (see Fig. \ref{fig_strf} (c)), while the Cu atoms do not contribute since their structure factor vanishes (Fig. \ref{fig_strf} (b)), similarly to the $k = 0$ case; \label{enum_1}
    \item ME toroidal moments. In contrast with the $k = 0$ case, the scattering wave vector does not lie in the a-c plane, hence the $\widetilde{\mathbf{t}} \wedge \mathbf{q}$ term in Eq. \eqref{eq6} displays a non-zero component perpendicular to the scattering plane; \label{enum_2}
    \item ME quadrupoles. In contrast with the $k = 0$ case, the $q_{xz}$ and $q_{z^2}$ components, which contribute to $M_{z} (\mk)$, have a non-zero structure factor. \label{enum_3}
\end{enumerate}

In Fig. \ref{fig_Pyy_Pzz} (b) we consider $k = 1$, i.e. $\mathbf{q} = (-1/2, 1, 1/2)$, and report the experimental results, which show that $|P_{yy}|, |P_{zz}| < |P_{xx}|$, in agreement with the theoretical arguments discussed above, and we compare them with the DFT predictions. In the same panel, we show also the DFT results decomposed to include separately the different sources, \ref{enum_1}-\ref{enum_3}, listed above. Remarkably, the non-collinear magnetic dipolar components play a marginal role. On the other hand, we demonstrate that the ME quadrupole moments are primarily responsible for the reduction of $P_{yy}$ and $P_{zz}$. 

This reduction in $P_{yy}$ and $P_{zz}$ represents the first direct experimental proof compatible with spontaneous long-ranged order of inversion-symmetry-breaking local ME multipoles. The excellent agreement between the SNP measurements of CuO and the DFT calculated scattering cross-sections lend support to our case that the ME multipoles arising from the oxygen ligands playing a predominant role in the additional neutron scattering processes, which has not been reported before. 
To obtain an independent measurement of the ME multipoles in CuO, we also measured the diagonal matrix elements of the $\textbf{q}=(-1/2,1,1/2)$ reflection on a different instrument (TASP) with a different implementation of SNP (MuPAD), as described earlier. The measured polarization matrix elements are $|P_{xx}| = 1.000 \pm 0.012$, $|P_{yy}| = 0.979 \pm 0.012$, $|P_{zz}| = 0.969 \pm 0.012$. Despite the significantly lower neutron flux on TASP compared to that on the D3 instrument, both measurements are consistent with each other. 

\section{Summary and Outlook} 
\label{final} 

In summary, in this work we reviewed the theoretical formulation of magnetic neutron diffraction, with particular emphasis on the contribution of the ME multipoles to the magnetic scattering cross-section. We demonstrated how to connect DFT predictions of the size of the ME multipoles to computations of the magnetic and ME scattering intensity, which is a directly measurable quantity. Additionally, we proposed ways of detecting signatures of ME multipoles, both with unpolarized and polarized neutrons.

As a case study, we discussed in detail the presence of ME multipoles in CuO and their effect on neutron diffraction. Our first-principles calculations showed that Cu ions support an anti-ferroic arrangement of toroidal moments in the $\mathbf{a}$-$\mathbf{c}$ plane (in agreement with previous predictions \cite{Lovesey_multipoles_1} based on model wave functions) as well as ME quadrupoles which had not been discussed previously. We found that the O ions carry ME multipoles that are approximately two orders of magnitude larger than those of Cu, indicating that the previously neglected O multipolar contributions are essential to understanding the details of the magnetic order in CuO. Based on these results, we analyzed the diagonal entries of the neutron polarization matrix for the reflections $\mathbf{q} = (-1/2, 0, 1/2)$ and $\mathbf{q} = (-1/2, 1, 1/2)$ and compared them to our SNP experimental data. At $\mathbf{q} = (-1/2, 1, 1/2)$ the measurements show a reduction of the $P_{yy}$ and $P_{zz}$ components compared to $P_{xx}$, consistent with the presence of ME multipoles predicted by DFT, and mostly caused by the quadrupoles on the O sites. Our calculations show that the scattering amplitude at wave vectors $\mathbf{q}$ perpendicular to the $\mathbf{b}$ axis is not affected by the ME multipoles, thus explaining why previous measurements performed with the crystal aligned along $\mathbf{b}$ were not sensitive to the multipolar order~\cite{Babkevich_2012_CuO_SNP,Brown_1991_CuO_SNP,Ain_1992_CuO_SNP,Qureshi_2020_CuO_SNP}.

An important next step in the study of the long-ranged order of ME multipoles in the AF1 phase of CuO will be the careful measurement of the off-diagonal entries of the polarization matrices. Due to the small scattering cross-sections, such measurements would require higher-flux neutron sources (e.g. the European Spallation Source) to achieve data acquisition times that are reasonable. 

More broadly, there is a strong imperative to search for cleaner candidates that host long-ranged order of ME multipoles, in which the ME reflections are not contaminated either by structural or magnetic Bragg peaks. The workflow presented in this work goes beyond symmetry analysis and therefore lays the essential ground work for the search for more promising candidate materials. 

\section*{Acknowledgements}
We thank Andrew T. Boothroyd, Andrew Wildes and Niels Bech Christensen for many fruitful discussions, Sebastian Vial, J. Alberto Rodr\'iguez-Velamaz\'an and Alexandra Turrini for technical help, Stephen Lovesey for drawing our attention to CuO and Vassil Skumryev and Marin Gospodinov for the provision of the CuO crystals. This work was funded by the European Research Council under the European Union’s Horizon 2020 research and innovation program synergy grant (HERO, Grant No. 810451). Computational resources were provided by ETH Z\"urich’s Euler cluster. JRS acknowledges support from the Singapore National Science Scholarship from the Agency for Science Technology and Research. The spherical neutron polarimetry proposal numbers are 5-41-1172 (ILL) and 20210052 (SINQ). The D3 data is available at~\cite{ILL_data}.
\appendix 
\vspace{1cm}


\section{Irreducible spherical magnetic and ME tensors}
\label{app_tensor_multipoles}
\begingroup
\begin{table*}[t]
\begin{tabular}{|c|c|c|c|} 
\hline 
$L' = 0$ & \multicolumn{3}{c|}{$L' = 1$} \\
\hline
$M' = 0$ & $M' = -1$ & $M' = 0$ & $M' = 1$ \\
\hline 
$\langle 1 {-1} \, 1 1 \lvert 0 0 \rangle = 1/\sqrt{3}$ & $\langle 1 {-1} \, 1 0 \lvert 1 {-1} \rangle = -1/\sqrt{2}$ & $\langle 1 1 \, 1 {-1} \lvert 1 0 \rangle = 1/\sqrt{2}$ & $\langle 1 1 \, 1 0 \lvert 1 1 \rangle = 1/\sqrt{2}$ \\
\hline 
$\langle 1 0 \, 1 0 \lvert 0 0 \rangle = -1/\sqrt{3}$ & $\langle 1 0 \, 1 {-1} \lvert 1 {-1} \rangle = 1/\sqrt{2}$ & $\langle 1 0 \, 1 0 \lvert 1 0 \rangle = 0$ & $\langle 1 0 \, 1 1 \lvert 1 1 \rangle = -1/\sqrt{2}$ \\
\hline 
$\langle 1 1 \, 1 {-1} \lvert 0 0 \rangle = 1/\sqrt{3}$ & & $\langle 1 {-1} \, 1 1 \lvert 1 0 \rangle = -1/\sqrt{2}$ & \\
\hline
\end{tabular}
\begin{tabular}{|c|c|c|c|c|}
\hline 
\multicolumn{5}{|c|}{$L' = 2$} \\
\hline
$M' = -2$ & $M' = -1$ & $M' = 0$ & $M' = 1$ & $M' = 2$\\
\hline 
$\langle 1 {-1} \, 1 {-1} \lvert 2 {-2} \rangle = 1$ & $\langle 1 0 \, 1 {-1} \lvert 2 {-1} \rangle = 1/\sqrt{2}$ & $\langle 1 {-1} \, 1 1 \lvert 2 0 \rangle = 1/\sqrt{6}$ & $\langle 1 0 \, 1 1 \lvert 2 1 \rangle = 1/\sqrt{2}$ & $\langle 1 1 \, 1 1 \lvert 2 2 \rangle = 1$ \\
\hline 
& $\langle 1 {-1} \, 1 0 \lvert 2 {-1} \rangle = 1/\sqrt{2}$ & $\langle 1 0 \, 1 0 \lvert 2 0 \rangle = \sqrt{2/3}$ & $\langle 1 1 \, 1 0 \lvert 2 1 \rangle = 1/\sqrt{2}$ & \\
\hline 
& & $\langle 1 1 \, 1 {-1} \lvert 2 0 \rangle = 1/\sqrt{6}$ & & \\
\hline
\end{tabular}
\caption{Clebsch-Gordan coefficients useful for evaluating the tensor moment $T^{(\text{s})}_{L' M'}$ (Eq. \eqref{eq4}) in the cases $L' = 0$, $L' = 1$, and $L' = 2$.}
\label{app_t1}
\end{table*}
\endgroup

In this Appendix we discuss Eq. \eqref{eq4} in more detail. In particular, we consider the cases $L = 0$ and $L = 1$ mentioned in Section \ref{spin_contribution}. 
\subsection{\texorpdfstring{$L = 0$}{L0}} In this case $\widebar{M} = 0$, hence: 
\begin{equation}
    T^{(\text{s})}_{L' M'} = \sqrt{4 \pi} \left[ \sum_{\bar{p}} s_{\bar{p}} \, Y_{00} (\hat{\mathbf{r}}) \langle 1 \bar{p} \, 0 0 \lvert L' M' \rangle \right]. 
\end{equation}
The Clebsch-Gordan coefficient is non-zero if $L' = 1$: in particular, we have $\langle 1 \bar{p} \, 0 0 \lvert 1 M' \rangle = \delta_{\bar{p} M'}$. Since $Y_{00} (\hat{\mathbf{r}}) = 1 / \sqrt{4 \pi}$, $T^{(\text{s})}_{L' M'}$ reads 
\begin{equation}
    T^{(\text{s})}_{1 M'}= s_{M'},
\end{equation}
hence $\mathbf{T}^{(\text{s})}_1 = \mathbf{s}$. We remark that the index $M' = (-L', -L'+1 \dots L-1, L)$ identifies the spherical components of the tensor. In this case, the spherical components $M' = (-1, 0, +1)$ of a vector $\mathbf{v}$ are connected to its cartesian components by the following relationships: 
\begin{align}
\label{app_eq_1}
    v_x &= \frac{1}{\sqrt{2}} \left( v_{-1} - v_1 \right), \\
    v_y &= \frac{i}{\sqrt{2}} \left( v_{-1} + v_1 \right), \\
    v_z &= v_0.
\label{app_eq_2}
\end{align}
\subsection{\texorpdfstring{$L = 1$}{L1}} In this case, $L'$ can take the values $L' = 0, 1, 2$. 
\paragraph{\texorpdfstring{$L' = 0$}{L'0}} We have: 
\begin{equation}
    T^{(\text{s})}_{0 0} = \sqrt{4 \pi} \left[ \sum_{\bar{p} \widebar{M}} s_{\bar{p}} \, Y_{1\widebar{M}} (\hat{\mathbf{r}}) \langle 1 \bar{p} \, 1 \widebar{M} \lvert 0 0 \rangle \right]. 
\end{equation}
The allowed non-zero Clebsch-Gordan coefficients are such that $\bar{p} + \widebar{M} = 0$; their values are reported in Table \ref{app_t1}. Since $Y_{1\widebar{M}} (\hat{\mathbf{r}}) = 1/2 \, \sqrt{3/\pi} \, \hat{r}_{\widebar{M}}$, where $\hat{r}_{\widebar{M}}$ is the $\widebar{M}$-th spherical component of $\hat{\mathbf{r}}$, after substituting the values of the Clebsch-Gordan coefficients and the expression for $Y_{1\widebar{M}} (\hat{\mathbf{r}})$, $T^{(\text{s})}_{0 0}$ reads
\begin{equation}
    T^{(\text{s})}_{00} = s_{-1} \hat{r}_1 - s_0 \hat{r}_0 + s_1 \hat{r}_{-1},
\end{equation}
which, after converting the spherical components into the cartesian ones using the inverse of Eqs. \eqref{app_eq_1}-\eqref{app_eq_2}, becomes 
\begin{equation}
    T^{(\text{s})}_{00} = - \hat{\mathbf{r}} \cdot \mathbf{s} = - \widetilde{a}.
\end{equation}
\paragraph{\texorpdfstring{$L' = 1$}{L'1}}
The expression for the tensor moment $T^{(\text{s})}_{1 M'}$, with $M' = (-1, 0, +1)$ in this case reads
\begin{equation}
    T^{(\text{s})}_{1 M'} = \sqrt{4 \pi} \left[ \sum_{\bar{p} \widebar{M}} s_{\bar{p}} \, Y_{1\widebar{M}} (\hat{\mathbf{r}}) \langle 1 \bar{p} \, 1 \widebar{M} \lvert 1 M' \rangle \right].
\end{equation}
The allowed Clebsch-Gordan coefficients in this case are the ones with $\bar{p} + \widebar{M} = M'$, as reported in Table \ref{app_t1}. After substituting the expression for $Y_{1\widebar{M}} (\hat{\mathbf{r}})$, the three spherical components of $\mathbf{T}^{(\text{s})}_1$ are
\begin{align}
    T^{(\text{s})}_{1 \, -1} &= \sqrt{\frac{3}{2}} \left( s_0 \hat{r}_{-1} - s_{-1} \hat{r}_0 \right), \\ 
    T^{(\text{s})}_{1 \, 0} &= \sqrt{\frac{3}{2}} \left( s_1 \hat{r}_{-1} - s_{-1} \hat{r}_1 \right) \text{and}\\ 
    T^{(\text{s})}_{1 \, 1} &= \sqrt{\frac{3}{2}} \left( s_1 \hat{r}_{0} - s_{0} \hat{r}_1 \right).
\end{align}
After transforming $\mathbf{s}$ and $\hat{\mathbf{r}}$ into the cartesian system, using Eqs. \eqref{app_eq_1}-\eqref{app_eq_2} it is possible to compute the cartesian components of $\mathbf{T}^{(\text{s})}_1$, which are given by
\begin{align}
    T^{(\text{s})}_{1 \, x} &= i \sqrt{\frac{3}{2}} \left( s_y \hat{z} - s_z \hat{y} \right), \\ 
    T^{(\text{s})}_{1 \, y} &= i \sqrt{\frac{3}{2}} \left( s_z \hat{x} - s_x \hat{z} \right) \text{and} \\
    T^{(\text{s})}_{1 \, z} &= i \sqrt{\frac{3}{2}} \left( s_x \hat{y} - s_y \hat{x} \right), 
\end{align}
hence 
\begin{equation}
    \mathbf{T}^{(\text{s})}_1 = - i \sqrt{\frac{3}{2}} \, (\hat{\mathbf{r}} \wedge \mathbf{s}) = - i \sqrt{\frac{3}{2}} \, \widetilde{\mathbf{t}},
\end{equation}
where $\widetilde{\mathbf{t}}$ is the toroidal moment operator.

\paragraph{\texorpdfstring{$L' = 2$}{L'2}} The resulting tensor moment $T^{(\text{s})}_{2M'}$, with $M'$ running from $-2$ to $+2$, following Eq. \eqref{eq4} is 
\begin{equation}
    T^{(\text{s})}_{1 M'} = \sqrt{4 \pi} \left[ \sum_{\bar{p} \widebar{M}} s_{\bar{p}} \, Y_{1\widebar{M}} (\hat{\mathbf{r}}) \langle 1 \bar{p} \, 1 \widebar{M} \lvert 2 M' \rangle \right]. 
\end{equation}
After inserting the expression for $Y_{1\widebar{M}}$ and the allowed Clebsch-Gordan coefficients, reported in Table \ref{app_t1}, the spherical components of $\mathbf{T}^{(\text{s})}_2$ read 
\begin{align}
    T^{(\text{s})}_{2 \, -2} &= \sqrt{3} \, s_{-1} \hat{r}_{-1}, \\
    T^{(\text{s})}_{2 \, -1} &= \sqrt{\frac{3}{2}} \left( s_{0} \hat{r}_{-1} + s_{-1} \hat{r}_{0} \right), \\
    T^{(\text{s})}_{2 \, 0} &= \frac{1}{\sqrt{2}} \left( s_{-1} \hat{r}_{1} + s_{1} \hat{r}_{-1} \right) + \sqrt{2} \, s_0 \hat{r}_0, \\
    T^{(\text{s})}_{2 \, 1} &= \sqrt{\frac{3}{2}} \left( s_{0} \hat{r}_{1} + s_{1} \hat{r}_{0} \right), \\
    T^{(\text{s})}_{2 \, 2} &= \sqrt{3} \, s_{1} \hat{r}_{1}.
\end{align}
By transforming $\mathbf{s}$ and $\hat{\mathbf{r}}$ into cartesian components, using the definition of the ME quadrupoles provided in Section \ref{theory}, and finally transforming the spherical tensor $\mathbf{T}^{(\text{s})}_2$ into cartesian components \cite{Bransden}, we have
\begin{align}
    T^{(\text{s})}_{2 \, yz} &= \frac{i}{\sqrt{2}} \left( T^{(\text{s})}_{2 \, -2} - T^{(\text{s})}_{2 \, 2} \right) = \sqrt{6} \widetilde{q}_{xy}, \\
    T^{(\text{s})}_{2 \, xy} &= \frac{i}{\sqrt{2}} \left( T^{(\text{s})}_{2 \, -1} + T^{(\text{s})}_{2 \, 1} \right) = \sqrt{6} \widetilde{q}_{yz}, \\ 
    T^{(\text{s})}_{2 \, z^2} &= T^{(\text{s})}_{2 \, 0} = \frac{3}{\sqrt{2}} \widetilde{q}_{z^2}, \\ 
    T^{(\text{s})}_{2 \, xz} &= \frac{1}{\sqrt{2}} \left( T^{(\text{s})}_{2 \, -1} - T^{(\text{s})}_{2 \, 1} \right) = \sqrt{6} \widetilde{q}_{xz}, \\
    T^{(\text{s})}_{2 \, x^2-y^2} &= \frac{1}{\sqrt{2}} \left( T^{(\text{s})}_{2 \, -2} + T^{(\text{s})}_{2 \, 2} \right) = \sqrt{\frac{3}{2}} \widetilde{q}_{xy}. 
\end{align}
\bibliographystyle{apsrev4-2}
\bibliography{references,references_2}

\begin{thebibliography}{50}%
\makeatletter
\providecommand \@ifxundefined [1]{%
 \@ifx{#1\undefined}
}%
\providecommand \@ifnum [1]{%
 \ifnum #1\expandafter \@firstoftwo
 \else \expandafter \@secondoftwo
 \fi
}%
\providecommand \@ifx [1]{%
 \ifx #1\expandafter \@firstoftwo
 \else \expandafter \@secondoftwo
 \fi
}%
\providecommand \natexlab [1]{#1}%
\providecommand \enquote  [1]{``#1''}%
\providecommand \bibnamefont  [1]{#1}%
\providecommand \bibfnamefont [1]{#1}%
\providecommand \citenamefont [1]{#1}%
\providecommand \href@noop [0]{\@secondoftwo}%
\providecommand \href [0]{\begingroup \@sanitize@url \@href}%
\providecommand \@href[1]{\@@startlink{#1}\@@href}%
\providecommand \@@href[1]{\endgroup#1\@@endlink}%
\providecommand \@sanitize@url [0]{\catcode `\\12\catcode `\$12\catcode
  `\&12\catcode `\#12\catcode `\^12\catcode `\_12\catcode `\%12\relax}%
\providecommand \@@startlink[1]{}%
\providecommand \@@endlink[0]{}%
\providecommand \url  [0]{\begingroup\@sanitize@url \@url }%
\providecommand \@url [1]{\endgroup\@href {#1}{\urlprefix }}%
\providecommand \urlprefix  [0]{URL }%
\providecommand \Eprint [0]{\href }%
\providecommand \doibase [0]{https://doi.org/}%
\providecommand \selectlanguage [0]{\@gobble}%
\providecommand \bibinfo  [0]{\@secondoftwo}%
\providecommand \bibfield  [0]{\@secondoftwo}%
\providecommand \translation [1]{[#1]}%
\providecommand \BibitemOpen [0]{}%
\providecommand \bibitemStop [0]{}%
\providecommand \bibitemNoStop [0]{.\EOS\space}%
\providecommand \EOS [0]{\spacefactor3000\relax}%
\providecommand \BibitemShut  [1]{\csname bibitem#1\endcsname}%
\let\auto@bib@innerbib\@empty
\bibitem [{\citenamefont {Boothroyd}(2020)}]{Boothroyd_book}%
  \BibitemOpen
  \bibfield  {author} {\bibinfo {author} {\bibfnamefont {A.~T.}\ \bibnamefont
  {Boothroyd}},\ }\href {https://doi.org/10.1093/oso/9780198862314.001.0001}
  {\emph {\bibinfo {title} {Principles of Neutron Scattering from Condensed
  Matter}}}\ (\bibinfo  {publisher} {Oxford University Press},\ \bibinfo {year}
  {2020})\BibitemShut {NoStop}%
\bibitem [{\citenamefont {Spaldin}\ \emph {et~al.}(2013)\citenamefont
  {Spaldin}, \citenamefont {Fechner}, \citenamefont {Bousquet}, \citenamefont
  {Balatsky},\ and\ \citenamefont {Nordstr{\"{o}}m}}]{me_multipoles}%
  \BibitemOpen
  \bibfield  {author} {\bibinfo {author} {\bibfnamefont {N.~A.}\ \bibnamefont
  {Spaldin}}, \bibinfo {author} {\bibfnamefont {M.}~\bibnamefont {Fechner}},
  \bibinfo {author} {\bibfnamefont {E.}~\bibnamefont {Bousquet}}, \bibinfo
  {author} {\bibfnamefont {A.}~\bibnamefont {Balatsky}},\ and\ \bibinfo
  {author} {\bibfnamefont {L.}~\bibnamefont {Nordstr{\"{o}}m}},\ }\href
  {https://doi.org/10.1103/PhysRevB.88.094429} {\bibfield  {journal} {\bibinfo
  {journal} {Phys. Rev. B}\ }\textbf {\bibinfo {volume} {88}},\ \bibinfo
  {pages} {094429} (\bibinfo {year} {2013})}\BibitemShut {NoStop}%
\bibitem [{\citenamefont {Blume}(1963)}]{Blume}%
  \BibitemOpen
  \bibfield  {author} {\bibinfo {author} {\bibfnamefont {M.}~\bibnamefont
  {Blume}},\ }\href {https://doi.org/10.1103/PhysRev.130.1670} {\bibfield
  {journal} {\bibinfo  {journal} {Phys. Rev.}\ }\textbf {\bibinfo {volume}
  {130}},\ \bibinfo {pages} {1670} (\bibinfo {year} {1963})}\BibitemShut
  {NoStop}%
\bibitem [{\citenamefont {Johnston}(1966)}]{Johnston_orbital}%
  \BibitemOpen
  \bibfield  {author} {\bibinfo {author} {\bibfnamefont {D.~F.}\ \bibnamefont
  {Johnston}},\ }\href {https://doi.org/10.1088/0370-1328/88/1/305} {\bibfield
  {journal} {\bibinfo  {journal} {Proceedings of the Physical Society}\
  }\textbf {\bibinfo {volume} {88}},\ \bibinfo {pages} {37} (\bibinfo {year}
  {1966})}\BibitemShut {NoStop}%
\bibitem [{\citenamefont {Lovesey}\ and\ \citenamefont
  {Rimmer}(1969)}]{Lovesey_review_multipoles}%
  \BibitemOpen
  \bibfield  {author} {\bibinfo {author} {\bibfnamefont {S.~W.}\ \bibnamefont
  {Lovesey}}\ and\ \bibinfo {author} {\bibfnamefont {D.~E.}\ \bibnamefont
  {Rimmer}},\ }\href {https://doi.org/10.1088/0034-4885/32/1/307} {\bibfield
  {journal} {\bibinfo  {journal} {Reports on Progress in Physics}\ }\textbf
  {\bibinfo {volume} {32}},\ \bibinfo {pages} {333} (\bibinfo {year}
  {1969})}\BibitemShut {NoStop}%
\bibitem [{\citenamefont {Lovesey}(1969)}]{Lovesey_multipoles_note}%
  \BibitemOpen
  \bibfield  {author} {\bibinfo {author} {\bibfnamefont {S.~W.}\ \bibnamefont
  {Lovesey}},\ }\href {https://doi.org/10.1088/0022-3719/2/3/311} {\bibfield
  {journal} {\bibinfo  {journal} {Journal of Physics C: Solid State Physics}\
  }\textbf {\bibinfo {volume} {2}},\ \bibinfo {pages} {470} (\bibinfo {year}
  {1969})}\BibitemShut {NoStop}%
\bibitem [{\citenamefont {Lovesey}(1984)}]{Lovesey_book_right}%
  \BibitemOpen
  \bibfield  {author} {\bibinfo {author} {\bibfnamefont {S.~W.}\ \bibnamefont
  {Lovesey}},\ }\href@noop {} {\emph {\bibinfo {title} {Theory of Neutron
  Scattering from Condensed Matter}}}\ (\bibinfo  {publisher} {Oxford
  University Press},\ \bibinfo {year} {1984})\BibitemShut {NoStop}%
\bibitem [{\citenamefont {Scagnoli}\ \emph {et~al.}(2011)\citenamefont
  {Scagnoli}, \citenamefont {Staub}, \citenamefont {Bodenthin}, \citenamefont
  {De~Souza}, \citenamefont {Garc{\'{i}}a-Fern{\'{a}}ndez}, \citenamefont
  {Garganourakis}, \citenamefont {Boothroyd}, \citenamefont {Prabhakaran},\
  and\ \citenamefont {Lovesey}}]{toroidal_CuO}%
  \BibitemOpen
  \bibfield  {author} {\bibinfo {author} {\bibfnamefont {V.}~\bibnamefont
  {Scagnoli}}, \bibinfo {author} {\bibfnamefont {U.}~\bibnamefont {Staub}},
  \bibinfo {author} {\bibfnamefont {Y.}~\bibnamefont {Bodenthin}}, \bibinfo
  {author} {\bibfnamefont {R.~A.}\ \bibnamefont {De~Souza}}, \bibinfo {author}
  {\bibfnamefont {M.}~\bibnamefont {Garc{\'{i}}a-Fern{\'{a}}ndez}}, \bibinfo
  {author} {\bibfnamefont {M.}~\bibnamefont {Garganourakis}}, \bibinfo {author}
  {\bibfnamefont {A.~T.}\ \bibnamefont {Boothroyd}}, \bibinfo {author}
  {\bibfnamefont {D.}~\bibnamefont {Prabhakaran}},\ and\ \bibinfo {author}
  {\bibfnamefont {S.~W.}\ \bibnamefont {Lovesey}},\ }\href
  {https://doi.org/10.1126/science.1201061} {\bibfield  {journal} {\bibinfo
  {journal} {Science}\ }\textbf {\bibinfo {volume} {332}},\ \bibinfo {pages}
  {696} (\bibinfo {year} {2011})}\BibitemShut {NoStop}%
\bibitem [{\citenamefont {Lovesey}(2014)}]{Lovesey_multipoles_1}%
  \BibitemOpen
  \bibfield  {author} {\bibinfo {author} {\bibfnamefont {S.~W.}\ \bibnamefont
  {Lovesey}},\ }\href {https://doi.org/10.1088/0953-8984/26/35/356001}
  {\bibfield  {journal} {\bibinfo  {journal} {J. Phys.: Condens. Matter}\
  }\textbf {\bibinfo {volume} {26}},\ \bibinfo {pages} {356001} (\bibinfo
  {year} {2014})}\BibitemShut {NoStop}%
\bibitem [{\citenamefont {Lovesey}(2015)}]{Lovesey_multipoles_2}%
  \BibitemOpen
  \bibfield  {author} {\bibinfo {author} {\bibfnamefont {S.~W.}\ \bibnamefont
  {Lovesey}},\ }\href {https://doi.org/10.1088/0031-8949/90/10/108011}
  {\bibfield  {journal} {\bibinfo  {journal} {Physica Scripta}\ }\textbf
  {\bibinfo {volume} {90}},\ \bibinfo {pages} {108011} (\bibinfo {year}
  {2015})}\BibitemShut {NoStop}%
\bibitem [{\citenamefont {Lovesey}\ \emph {et~al.}(2015)\citenamefont
  {Lovesey}, \citenamefont {Khalyavin},\ and\ \citenamefont {Van
  Der~Laan}}]{Lovesey_multipoles_4}%
  \BibitemOpen
  \bibfield  {author} {\bibinfo {author} {\bibfnamefont {S.~W.}\ \bibnamefont
  {Lovesey}}, \bibinfo {author} {\bibfnamefont {D.~D.}\ \bibnamefont
  {Khalyavin}},\ and\ \bibinfo {author} {\bibfnamefont {G.}~\bibnamefont {Van
  Der~Laan}},\ }\href {https://doi.org/10.1088/0031-8949/91/1/015803}
  {\bibfield  {journal} {\bibinfo  {journal} {Physica Scripta}\ }\textbf
  {\bibinfo {volume} {91}},\ \bibinfo {pages} {015803} (\bibinfo {year}
  {2015})}\BibitemShut {NoStop}%
\bibitem [{\citenamefont {Lovesey}\ and\ \citenamefont
  {Khalyavin}(2017{\natexlab{a}})}]{Lovesey_multipoles_3}%
  \BibitemOpen
  \bibfield  {author} {\bibinfo {author} {\bibfnamefont {S.~W.}\ \bibnamefont
  {Lovesey}}\ and\ \bibinfo {author} {\bibfnamefont {D.~D.}\ \bibnamefont
  {Khalyavin}},\ }\href {https://doi.org/10.1088/1361-648X/aa5ad8} {\bibfield
  {journal} {\bibinfo  {journal} {Journal of Physics Condensed Matter}\
  }\textbf {\bibinfo {volume} {29}},\ \bibinfo {pages} {215603} (\bibinfo
  {year} {2017}{\natexlab{a}})}\BibitemShut {NoStop}%
\bibitem [{\citenamefont {Lovesey}\ and\ \citenamefont
  {Khalyavin}(2017{\natexlab{b}})}]{Lovesey_multipoles_5}%
  \BibitemOpen
  \bibfield  {author} {\bibinfo {author} {\bibfnamefont {S.~W.}\ \bibnamefont
  {Lovesey}}\ and\ \bibinfo {author} {\bibfnamefont {D.~D.}\ \bibnamefont
  {Khalyavin}},\ }\href {https://doi.org/10.1088/1361-648X/aa860f} {\bibfield
  {journal} {\bibinfo  {journal} {Journal of Physics Condensed Matter}\
  }\textbf {\bibinfo {volume} {29}},\ \bibinfo {pages} {455604} (\bibinfo
  {year} {2017}{\natexlab{b}})}\BibitemShut {NoStop}%
\bibitem [{\citenamefont {Lovesey}\ and\ \citenamefont
  {Khalyavin}(2018)}]{Lovesey_multipoles_pnictides}%
  \BibitemOpen
  \bibfield  {author} {\bibinfo {author} {\bibfnamefont {S.~W.}\ \bibnamefont
  {Lovesey}}\ and\ \bibinfo {author} {\bibfnamefont {D.~D.}\ \bibnamefont
  {Khalyavin}},\ }\href {https://doi.org/10.1103/PhysRevB.98.054434} {\bibfield
   {journal} {\bibinfo  {journal} {Physical Review B}\ }\textbf {\bibinfo
  {volume} {98}},\ \bibinfo {pages} {054434} (\bibinfo {year}
  {2018})}\BibitemShut {NoStop}%
\bibitem [{\citenamefont {Lovesey}\ \emph {et~al.}(2019)\citenamefont
  {Lovesey}, \citenamefont {Chatterji}, \citenamefont {Stunault}, \citenamefont
  {Khalyavin},\ and\ \citenamefont {Van Der~Laan}}]{Lovesey_anapoles}%
  \BibitemOpen
  \bibfield  {author} {\bibinfo {author} {\bibfnamefont {S.~W.}\ \bibnamefont
  {Lovesey}}, \bibinfo {author} {\bibfnamefont {T.}~\bibnamefont {Chatterji}},
  \bibinfo {author} {\bibfnamefont {A.}~\bibnamefont {Stunault}}, \bibinfo
  {author} {\bibfnamefont {D.~D.}\ \bibnamefont {Khalyavin}},\ and\ \bibinfo
  {author} {\bibfnamefont {G.}~\bibnamefont {Van Der~Laan}},\ }\href
  {https://doi.org/10.1103/PhysRevLett.122.047203} {\bibfield  {journal}
  {\bibinfo  {journal} {Physical Review Letters}\ }\textbf {\bibinfo {volume}
  {122}},\ \bibinfo {pages} {047203} (\bibinfo {year} {2019})}\BibitemShut
  {NoStop}%
\bibitem [{\citenamefont {Hohenberg}\ and\ \citenamefont
  {Kohn}(1964)}]{DFT_HK}%
  \BibitemOpen
  \bibfield  {author} {\bibinfo {author} {\bibfnamefont {P.}~\bibnamefont
  {Hohenberg}}\ and\ \bibinfo {author} {\bibfnamefont {W.}~\bibnamefont
  {Kohn}},\ }\href {https://doi.org/10.1103/PhysRev.136.B864} {\bibfield
  {journal} {\bibinfo  {journal} {Physical Review}\ }\textbf {\bibinfo {volume}
  {136}},\ \bibinfo {pages} {B864} (\bibinfo {year} {1964})}\BibitemShut
  {NoStop}%
\bibitem [{\citenamefont {Kohn}\ and\ \citenamefont {Sham}(1965)}]{DFT_KS}%
  \BibitemOpen
  \bibfield  {author} {\bibinfo {author} {\bibfnamefont {W.}~\bibnamefont
  {Kohn}}\ and\ \bibinfo {author} {\bibfnamefont {L.~J.}\ \bibnamefont
  {Sham}},\ }\href {https://doi.org/10.1103/PhysRev.140.A1133} {\bibfield
  {journal} {\bibinfo  {journal} {Physical Review}\ }\textbf {\bibinfo {volume}
  {140}},\ \bibinfo {pages} {A1133} (\bibinfo {year} {1965})}\BibitemShut
  {NoStop}%
\bibitem [{Elk()}]{Elk_code}%
  \BibitemOpen
  \href@noop {} {\bibinfo {title} {See
  \url{https://elk.sourceforge.io/}}}\BibitemShut {NoStop}%
\bibitem [{\citenamefont {Bultmark}\ \emph {et~al.}(2009)\citenamefont
  {Bultmark}, \citenamefont {Cricchio}, \citenamefont {Gr{\aa}n{\"{a}}s},\ and\
  \citenamefont {Nordstr{\"{o}}m}}]{multipole_decomposition}%
  \BibitemOpen
  \bibfield  {author} {\bibinfo {author} {\bibfnamefont {F.}~\bibnamefont
  {Bultmark}}, \bibinfo {author} {\bibfnamefont {F.}~\bibnamefont {Cricchio}},
  \bibinfo {author} {\bibfnamefont {O.}~\bibnamefont {Gr{\aa}n{\"{a}}s}},\ and\
  \bibinfo {author} {\bibfnamefont {L.}~\bibnamefont {Nordstr{\"{o}}m}},\
  }\href {https://doi.org/10.1103/PhysRevB.80.035121} {\bibfield  {journal}
  {\bibinfo  {journal} {Phys. Rev. B}\ }\textbf {\bibinfo {volume} {80}},\
  \bibinfo {pages} {035121} (\bibinfo {year} {2009})}\BibitemShut {NoStop}%
\bibitem [{\citenamefont {Liechtenstein}\ \emph {et~al.}(1995)\citenamefont
  {Liechtenstein}, \citenamefont {Anisimov},\ and\ \citenamefont
  {Zaanen}}]{Liechtenstein_U}%
  \BibitemOpen
  \bibfield  {author} {\bibinfo {author} {\bibfnamefont {A.~I.}\ \bibnamefont
  {Liechtenstein}}, \bibinfo {author} {\bibfnamefont {V.~I.}\ \bibnamefont
  {Anisimov}},\ and\ \bibinfo {author} {\bibfnamefont {J.}~\bibnamefont
  {Zaanen}},\ }\href {https://doi.org/10.1103/PhysRevB.52.R5467} {\bibfield
  {journal} {\bibinfo  {journal} {Phys. Rev. B}\ }\textbf {\bibinfo {volume}
  {52}},\ \bibinfo {pages} {R5467} (\bibinfo {year} {1995})}\BibitemShut
  {NoStop}%
\bibitem [{\citenamefont {Dudarev}\ and\ \citenamefont
  {Botton}(1998)}]{Dudarev_U}%
  \BibitemOpen
  \bibfield  {author} {\bibinfo {author} {\bibfnamefont {S.}~\bibnamefont
  {Dudarev}}\ and\ \bibinfo {author} {\bibfnamefont {G.}~\bibnamefont
  {Botton}},\ }\href {https://doi.org/10.1103/PhysRevB.57.1505} {\bibfield
  {journal} {\bibinfo  {journal} {Phys. Rev. B}\ }\textbf {\bibinfo {volume}
  {57}},\ \bibinfo {pages} {1505} (\bibinfo {year} {1998})}\BibitemShut
  {NoStop}%
\bibitem [{\citenamefont {Cococcioni}\ and\ \citenamefont
  {De~Gironcoli}(2005)}]{U_cococcioni}%
  \BibitemOpen
  \bibfield  {author} {\bibinfo {author} {\bibfnamefont {M.}~\bibnamefont
  {Cococcioni}}\ and\ \bibinfo {author} {\bibfnamefont {S.}~\bibnamefont
  {De~Gironcoli}},\ }\href {https://doi.org/10.1103/PhysRevB.71.035105}
  {\bibfield  {journal} {\bibinfo  {journal} {Phys. Rev. B}\ }\textbf {\bibinfo
  {volume} {71}},\ \bibinfo {pages} {035105} (\bibinfo {year}
  {2005})}\BibitemShut {NoStop}%
\bibitem [{\citenamefont {Urru}\ and\ \citenamefont {Spaldin}(2022)}]{mag_oct}%
  \BibitemOpen
  \bibfield  {author} {\bibinfo {author} {\bibfnamefont {A.}~\bibnamefont
  {Urru}}\ and\ \bibinfo {author} {\bibfnamefont {N.~A.}\ \bibnamefont
  {Spaldin}},\ }\href
  {https://doi.org/https://doi.org/10.1016/j.aop.2022.168964} {\bibfield
  {journal} {\bibinfo  {journal} {Annals of Physics}\ ,\ \bibinfo {pages}
  {168964}} (\bibinfo {year} {2022})}\BibitemShut {NoStop}%
\bibitem [{\citenamefont {Van Der~Laan}(1999)}]{van_der_laan_normalization}%
  \BibitemOpen
  \bibfield  {author} {\bibinfo {author} {\bibfnamefont {G.}~\bibnamefont {Van
  Der~Laan}},\ }\href {https://doi.org/10.1016/s0368-2048(98)00400-9}
  {\bibfield  {journal} {\bibinfo  {journal} {Journ. Electr. Spectr. Related
  Phen.}\ }\textbf {\bibinfo {volume} {101-103}},\ \bibinfo {pages} {859}
  (\bibinfo {year} {1999})}\BibitemShut {NoStop}%
\bibitem [{\citenamefont {Thole}\ and\ \citenamefont {Van
  Der~Laan}(1994)}]{van_der_laan_3}%
  \BibitemOpen
  \bibfield  {author} {\bibinfo {author} {\bibfnamefont {B.~T.}\ \bibnamefont
  {Thole}}\ and\ \bibinfo {author} {\bibfnamefont {G.}~\bibnamefont {Van
  Der~Laan}},\ }\href {https://doi.org/10.1103/PhysRevB.49.9613} {\bibfield
  {journal} {\bibinfo  {journal} {Phys. Rev. B}\ }\textbf {\bibinfo {volume}
  {49}},\ \bibinfo {pages} {9613} (\bibinfo {year} {1994})}\BibitemShut
  {NoStop}%
\bibitem [{\citenamefont {Varshalovich}\ \emph {et~al.}(1988)\citenamefont
  {Varshalovich}, \citenamefont {Moskalev},\ and\ \citenamefont
  {Khersonskii}}]{ang_mom_theory}%
  \BibitemOpen
  \bibfield  {author} {\bibinfo {author} {\bibfnamefont {D.~A.}\ \bibnamefont
  {Varshalovich}}, \bibinfo {author} {\bibfnamefont {A.~N.}\ \bibnamefont
  {Moskalev}},\ and\ \bibinfo {author} {\bibfnamefont {V.~K.}\ \bibnamefont
  {Khersonskii}},\ }\href {https://doi.org/10.1142/0270} {\emph {\bibinfo
  {title} {Quantum Theory of Angular Momentum}}}\ (\bibinfo  {publisher} {World
  Scientific Publishing Company},\ \bibinfo {year} {1988})\BibitemShut
  {NoStop}%
\bibitem [{\citenamefont {Qureshi}\ \emph {et~al.}(2020)\citenamefont
  {Qureshi}, \citenamefont {Ressouche}, \citenamefont {Mukhin}, \citenamefont
  {Gospodinov},\ and\ \citenamefont {Skumryev}}]{Qureshi_2020_CuO_SNP}%
  \BibitemOpen
  \bibfield  {author} {\bibinfo {author} {\bibfnamefont {N.}~\bibnamefont
  {Qureshi}}, \bibinfo {author} {\bibfnamefont {E.}~\bibnamefont {Ressouche}},
  \bibinfo {author} {\bibfnamefont {A.}~\bibnamefont {Mukhin}}, \bibinfo
  {author} {\bibfnamefont {M.}~\bibnamefont {Gospodinov}},\ and\ \bibinfo
  {author} {\bibfnamefont {V.}~\bibnamefont {Skumryev}},\ }\href
  {https://doi.org/10.1126/sciadv.aay7661} {\bibfield  {journal} {\bibinfo
  {journal} {Sci. Adv.}\ }\textbf {\bibinfo {volume} {6}},\ \bibinfo {pages}
  {aay7661} (\bibinfo {year} {2020})}\BibitemShut {NoStop}%
\bibitem [{\citenamefont {Joly}\ \emph {et~al.}(2012)\citenamefont {Joly},
  \citenamefont {Collins}, \citenamefont {Grenier}, \citenamefont {Tolentino},\
  and\ \citenamefont {De~Santis}}]{PhysRevB.86.220101}%
  \BibitemOpen
  \bibfield  {author} {\bibinfo {author} {\bibfnamefont {Y.}~\bibnamefont
  {Joly}}, \bibinfo {author} {\bibfnamefont {S.~P.}\ \bibnamefont {Collins}},
  \bibinfo {author} {\bibfnamefont {S.}~\bibnamefont {Grenier}}, \bibinfo
  {author} {\bibfnamefont {H.~C.~N.}\ \bibnamefont {Tolentino}},\ and\ \bibinfo
  {author} {\bibfnamefont {M.}~\bibnamefont {De~Santis}},\ }\href
  {https://doi.org/10.1103/PhysRevB.86.220101} {\bibfield  {journal} {\bibinfo
  {journal} {Phys. Rev. B}\ }\textbf {\bibinfo {volume} {86}},\ \bibinfo
  {pages} {220101} (\bibinfo {year} {2012})}\BibitemShut {NoStop}%
\bibitem [{\citenamefont {Giannozzi}\ \emph {et~al.}(2009)\citenamefont
  {Giannozzi}, \citenamefont {Baroni}, \citenamefont {Bonini}, \citenamefont
  {Calandra}, \citenamefont {Car}, \citenamefont {Cavazzoni}, \citenamefont
  {Ceresoli}, \citenamefont {Chiarotti}, \citenamefont {Cococcioni},
  \citenamefont {Dabo}, \citenamefont {Dal~Corso}, \citenamefont
  {De~Gironcoli}, \citenamefont {Fabris}, \citenamefont {Fratesi},
  \citenamefont {Gebauer}, \citenamefont {Gerstmann}, \citenamefont
  {Gougoussis}, \citenamefont {Kokalj}, \citenamefont {Lazzeri}, \citenamefont
  {Martin-Samos}, \citenamefont {Marzari}, \citenamefont {Mauri}, \citenamefont
  {Mazzarello}, \citenamefont {Paolini}, \citenamefont {Pasquarello},
  \citenamefont {Paulatto}, \citenamefont {Sbraccia}, \citenamefont {Scandolo},
  \citenamefont {Sclauzero}, \citenamefont {Seitsonen}, \citenamefont
  {Smogunov}, \citenamefont {Umari},\ and\ \citenamefont
  {Wentzcovitch}}]{QE_1}%
  \BibitemOpen
  \bibfield  {author} {\bibinfo {author} {\bibfnamefont {P.}~\bibnamefont
  {Giannozzi}}, \bibinfo {author} {\bibfnamefont {S.}~\bibnamefont {Baroni}},
  \bibinfo {author} {\bibfnamefont {N.}~\bibnamefont {Bonini}}, \bibinfo
  {author} {\bibfnamefont {M.}~\bibnamefont {Calandra}}, \bibinfo {author}
  {\bibfnamefont {R.}~\bibnamefont {Car}}, \bibinfo {author} {\bibfnamefont
  {C.}~\bibnamefont {Cavazzoni}}, \bibinfo {author} {\bibfnamefont
  {D.}~\bibnamefont {Ceresoli}}, \bibinfo {author} {\bibfnamefont {G.~L.}\
  \bibnamefont {Chiarotti}}, \bibinfo {author} {\bibfnamefont {M.}~\bibnamefont
  {Cococcioni}}, \bibinfo {author} {\bibfnamefont {I.}~\bibnamefont {Dabo}},
  \bibinfo {author} {\bibfnamefont {A.}~\bibnamefont {Dal~Corso}}, \bibinfo
  {author} {\bibfnamefont {S.}~\bibnamefont {De~Gironcoli}}, \bibinfo {author}
  {\bibfnamefont {S.}~\bibnamefont {Fabris}}, \bibinfo {author} {\bibfnamefont
  {G.}~\bibnamefont {Fratesi}}, \bibinfo {author} {\bibfnamefont
  {R.}~\bibnamefont {Gebauer}}, \bibinfo {author} {\bibfnamefont
  {U.}~\bibnamefont {Gerstmann}}, \bibinfo {author} {\bibfnamefont
  {C.}~\bibnamefont {Gougoussis}}, \bibinfo {author} {\bibfnamefont
  {A.}~\bibnamefont {Kokalj}}, \bibinfo {author} {\bibfnamefont
  {M.}~\bibnamefont {Lazzeri}}, \bibinfo {author} {\bibfnamefont
  {L.}~\bibnamefont {Martin-Samos}}, \bibinfo {author} {\bibfnamefont
  {N.}~\bibnamefont {Marzari}}, \bibinfo {author} {\bibfnamefont
  {F.}~\bibnamefont {Mauri}}, \bibinfo {author} {\bibfnamefont
  {R.}~\bibnamefont {Mazzarello}}, \bibinfo {author} {\bibfnamefont
  {S.}~\bibnamefont {Paolini}}, \bibinfo {author} {\bibfnamefont
  {A.}~\bibnamefont {Pasquarello}}, \bibinfo {author} {\bibfnamefont
  {L.}~\bibnamefont {Paulatto}}, \bibinfo {author} {\bibfnamefont
  {C.}~\bibnamefont {Sbraccia}}, \bibinfo {author} {\bibfnamefont
  {S.}~\bibnamefont {Scandolo}}, \bibinfo {author} {\bibfnamefont
  {G.}~\bibnamefont {Sclauzero}}, \bibinfo {author} {\bibfnamefont {A.~P.}\
  \bibnamefont {Seitsonen}}, \bibinfo {author} {\bibfnamefont {A.}~\bibnamefont
  {Smogunov}}, \bibinfo {author} {\bibfnamefont {P.}~\bibnamefont {Umari}},\
  and\ \bibinfo {author} {\bibfnamefont {R.~M.}\ \bibnamefont {Wentzcovitch}},\
  }\href {https://doi.org/10.1088/0953-8984/21/39/395502} {\bibfield  {journal}
  {\bibinfo  {journal} {J. Phys. Condens. Matter}\ }\textbf {\bibinfo {volume}
  {21}},\ \bibinfo {pages} {395502} (\bibinfo {year} {2009})}\BibitemShut
  {NoStop}%
\bibitem [{\citenamefont {Giannozzi}\ \emph {et~al.}(2017)\citenamefont
  {Giannozzi}, \citenamefont {Andreussi}, \citenamefont {Brumme}, \citenamefont
  {Bunau}, \citenamefont {Buongiorno~Nardelli}, \citenamefont {Calandra},
  \citenamefont {Car}, \citenamefont {Cavazzoni}, \citenamefont {Ceresoli},
  \citenamefont {Cococcioni}, \citenamefont {Colonna}, \citenamefont
  {Carnimeo}, \citenamefont {Dal~Corso}, \citenamefont {De~Gironcoli},
  \citenamefont {Delugas}, \citenamefont {Distasio}, \citenamefont {Ferretti},
  \citenamefont {Floris}, \citenamefont {Fratesi}, \citenamefont {Fugallo},
  \citenamefont {Gebauer}, \citenamefont {Gerstmann}, \citenamefont {Giustino},
  \citenamefont {Gorni}, \citenamefont {Jia}, \citenamefont {Kawamura},
  \citenamefont {Ko}, \citenamefont {Kokalj}, \citenamefont
  {K{\"{u}}c{\"{u}}kbenli}, \citenamefont {Lazzeri}, \citenamefont {Marsili},
  \citenamefont {Marzari}, \citenamefont {Mauri}, \citenamefont {Nguyen},
  \citenamefont {Nguyen}, \citenamefont {Otero-De-La-Roza}, \citenamefont
  {Paulatto}, \citenamefont {Ponc{\'{e}}}, \citenamefont {Rocca}, \citenamefont
  {Sabatini}, \citenamefont {Santra}, \citenamefont {Schlipf}, \citenamefont
  {Seitsonen}, \citenamefont {Smogunov}, \citenamefont {Timrov}, \citenamefont
  {Thonhauser}, \citenamefont {Umari}, \citenamefont {Vast}, \citenamefont
  {Wu},\ and\ \citenamefont {Baroni}}]{QE_2}%
  \BibitemOpen
  \bibfield  {author} {\bibinfo {author} {\bibfnamefont {P.}~\bibnamefont
  {Giannozzi}}, \bibinfo {author} {\bibfnamefont {O.}~\bibnamefont
  {Andreussi}}, \bibinfo {author} {\bibfnamefont {T.}~\bibnamefont {Brumme}},
  \bibinfo {author} {\bibfnamefont {O.}~\bibnamefont {Bunau}}, \bibinfo
  {author} {\bibfnamefont {M.}~\bibnamefont {Buongiorno~Nardelli}}, \bibinfo
  {author} {\bibfnamefont {M.}~\bibnamefont {Calandra}}, \bibinfo {author}
  {\bibfnamefont {R.}~\bibnamefont {Car}}, \bibinfo {author} {\bibfnamefont
  {C.}~\bibnamefont {Cavazzoni}}, \bibinfo {author} {\bibfnamefont
  {D.}~\bibnamefont {Ceresoli}}, \bibinfo {author} {\bibfnamefont
  {M.}~\bibnamefont {Cococcioni}}, \bibinfo {author} {\bibfnamefont
  {N.}~\bibnamefont {Colonna}}, \bibinfo {author} {\bibfnamefont
  {I.}~\bibnamefont {Carnimeo}}, \bibinfo {author} {\bibfnamefont
  {A.}~\bibnamefont {Dal~Corso}}, \bibinfo {author} {\bibfnamefont
  {S.}~\bibnamefont {De~Gironcoli}}, \bibinfo {author} {\bibfnamefont
  {P.}~\bibnamefont {Delugas}}, \bibinfo {author} {\bibfnamefont {R.~A.}\
  \bibnamefont {Distasio}}, \bibinfo {author} {\bibfnamefont {A.}~\bibnamefont
  {Ferretti}}, \bibinfo {author} {\bibfnamefont {A.}~\bibnamefont {Floris}},
  \bibinfo {author} {\bibfnamefont {G.}~\bibnamefont {Fratesi}}, \bibinfo
  {author} {\bibfnamefont {G.}~\bibnamefont {Fugallo}}, \bibinfo {author}
  {\bibfnamefont {R.}~\bibnamefont {Gebauer}}, \bibinfo {author} {\bibfnamefont
  {U.}~\bibnamefont {Gerstmann}}, \bibinfo {author} {\bibfnamefont
  {F.}~\bibnamefont {Giustino}}, \bibinfo {author} {\bibfnamefont
  {T.}~\bibnamefont {Gorni}}, \bibinfo {author} {\bibfnamefont
  {J.}~\bibnamefont {Jia}}, \bibinfo {author} {\bibfnamefont {M.}~\bibnamefont
  {Kawamura}}, \bibinfo {author} {\bibfnamefont {H.~Y.}\ \bibnamefont {Ko}},
  \bibinfo {author} {\bibfnamefont {A.}~\bibnamefont {Kokalj}}, \bibinfo
  {author} {\bibfnamefont {E.}~\bibnamefont {K{\"{u}}c{\"{u}}kbenli}}, \bibinfo
  {author} {\bibfnamefont {M.}~\bibnamefont {Lazzeri}}, \bibinfo {author}
  {\bibfnamefont {M.}~\bibnamefont {Marsili}}, \bibinfo {author} {\bibfnamefont
  {N.}~\bibnamefont {Marzari}}, \bibinfo {author} {\bibfnamefont
  {F.}~\bibnamefont {Mauri}}, \bibinfo {author} {\bibfnamefont {N.~L.}\
  \bibnamefont {Nguyen}}, \bibinfo {author} {\bibfnamefont {H.~V.}\
  \bibnamefont {Nguyen}}, \bibinfo {author} {\bibfnamefont {A.}~\bibnamefont
  {Otero-De-La-Roza}}, \bibinfo {author} {\bibfnamefont {L.}~\bibnamefont
  {Paulatto}}, \bibinfo {author} {\bibfnamefont {S.}~\bibnamefont
  {Ponc{\'{e}}}}, \bibinfo {author} {\bibfnamefont {D.}~\bibnamefont {Rocca}},
  \bibinfo {author} {\bibfnamefont {R.}~\bibnamefont {Sabatini}}, \bibinfo
  {author} {\bibfnamefont {B.}~\bibnamefont {Santra}}, \bibinfo {author}
  {\bibfnamefont {M.}~\bibnamefont {Schlipf}}, \bibinfo {author} {\bibfnamefont
  {A.~P.}\ \bibnamefont {Seitsonen}}, \bibinfo {author} {\bibfnamefont
  {A.}~\bibnamefont {Smogunov}}, \bibinfo {author} {\bibfnamefont
  {I.}~\bibnamefont {Timrov}}, \bibinfo {author} {\bibfnamefont
  {T.}~\bibnamefont {Thonhauser}}, \bibinfo {author} {\bibfnamefont
  {P.}~\bibnamefont {Umari}}, \bibinfo {author} {\bibfnamefont
  {N.}~\bibnamefont {Vast}}, \bibinfo {author} {\bibfnamefont {X.}~\bibnamefont
  {Wu}},\ and\ \bibinfo {author} {\bibfnamefont {S.}~\bibnamefont {Baroni}},\
  }\href {https://doi.org/10.1088/1361-648X/aa8f79} {\bibfield  {journal}
  {\bibinfo  {journal} {J. Phys. Condens. Matter}\ }\textbf {\bibinfo {volume}
  {29}},\ \bibinfo {pages} {465901} (\bibinfo {year} {2017})}\BibitemShut
  {NoStop}%
\bibitem [{the()}]{thermo_pw}%
  \BibitemOpen
  \href@noop {} {\bibinfo {title} {\texttt{thermo\_pw} is an extension of the
  \uppercase{Q}uantum \texttt{ESPRESSO} (\uppercase{QE}) package which provides
  an alternative organization of the \uppercase{QE} workflow for the most
  common tasks. \uppercase{F}or more information see
  \url{https://dalcorso.github.io/thermo_pw/}}}\BibitemShut {NoStop}%
\bibitem [{\citenamefont {Vanderbilt}(1990)}]{US_Vanderbilt}%
  \BibitemOpen
  \bibfield  {author} {\bibinfo {author} {\bibfnamefont {D.}~\bibnamefont
  {Vanderbilt}},\ }\href {https://doi.org/10.1103/PhysRevB.41.7892} {\bibfield
  {journal} {\bibinfo  {journal} {Phys. Rev. B}\ }\textbf {\bibinfo {volume}
  {41}},\ \bibinfo {pages} {7892} (\bibinfo {year} {1990})}\BibitemShut
  {NoStop}%
\bibitem [{\citenamefont {Dal~Corso}(2014)}]{pslibrary}%
  \BibitemOpen
  \bibfield  {author} {\bibinfo {author} {\bibfnamefont {A.}~\bibnamefont
  {Dal~Corso}},\ }\href {https://doi.org/10.1016/j.commatsci.2014.07.043}
  {\bibfield  {journal} {\bibinfo  {journal} {Comp. Mater. Sci.}\ }\textbf
  {\bibinfo {volume} {95}},\ \bibinfo {pages} {337} (\bibinfo {year}
  {2014})}\BibitemShut {NoStop}%
\bibitem [{psl()}]{pslibrary_2}%
  \BibitemOpen
  \href@noop {} {\bibinfo {title} {See
  \url{https://dalcorso.github.io/pslibrary/}}}\BibitemShut {NoStop}%
\bibitem [{\citenamefont {Himmetoglu}\ \emph {et~al.}(2011)\citenamefont
  {Himmetoglu}, \citenamefont {Wentzcovitch},\ and\ \citenamefont
  {Cococcioni}}]{CuO_cococcioni}%
  \BibitemOpen
  \bibfield  {author} {\bibinfo {author} {\bibfnamefont {B.}~\bibnamefont
  {Himmetoglu}}, \bibinfo {author} {\bibfnamefont {R.~M.}\ \bibnamefont
  {Wentzcovitch}},\ and\ \bibinfo {author} {\bibfnamefont {M.}~\bibnamefont
  {Cococcioni}},\ }\href {https://doi.org/10.1103/PhysRevB.84.115108}
  {\bibfield  {journal} {\bibinfo  {journal} {Phys. Rev. B}\ }\textbf {\bibinfo
  {volume} {84}},\ \bibinfo {pages} {115108} (\bibinfo {year}
  {2011})}\BibitemShut {NoStop}%
\bibitem [{\citenamefont {Asbrink}\ and\ \citenamefont
  {Waskowska}(1991)}]{CuO_structure}%
  \BibitemOpen
  \bibfield  {author} {\bibinfo {author} {\bibfnamefont {S.}~\bibnamefont
  {Asbrink}}\ and\ \bibinfo {author} {\bibfnamefont {A.}~\bibnamefont
  {Waskowska}},\ }\href {https://doi.org/10.1088/0953-8984/3/42/012} {\bibfield
   {journal} {\bibinfo  {journal} {Journal of Physics: Condensed Matter}\
  }\textbf {\bibinfo {volume} {3}},\ \bibinfo {pages} {8173} (\bibinfo {year}
  {1991})}\BibitemShut {NoStop}%
\bibitem [{\citenamefont {Monkhorst}\ and\ \citenamefont
  {Pack}(1976)}]{Monkhorst_Pack_mesh}%
  \BibitemOpen
  \bibfield  {author} {\bibinfo {author} {\bibfnamefont {H.~J.}\ \bibnamefont
  {Monkhorst}}\ and\ \bibinfo {author} {\bibfnamefont {J.~D.}\ \bibnamefont
  {Pack}},\ }\href {https://doi.org/10.1103/PhysRevB.13.5188} {\bibfield
  {journal} {\bibinfo  {journal} {Phys. Rev. B}\ }\textbf {\bibinfo {volume}
  {13}},\ \bibinfo {pages} {5188} (\bibinfo {year} {1976})}\BibitemShut
  {NoStop}%
\bibitem [{\citenamefont {Clementi}\ and\ \citenamefont
  {Roetti}(1974)}]{radial_functions}%
  \BibitemOpen
  \bibfield  {author} {\bibinfo {author} {\bibfnamefont {E.}~\bibnamefont
  {Clementi}}\ and\ \bibinfo {author} {\bibfnamefont {C.}~\bibnamefont
  {Roetti}},\ }\href {https://doi.org/10.1016/S0092-640X(74)80016-1} {\bibfield
   {journal} {\bibinfo  {journal} {Atomic Data and Nuclear Data Tables}\
  }\textbf {\bibinfo {volume} {14}},\ \bibinfo {pages} {177} (\bibinfo {year}
  {1974})}\BibitemShut {NoStop}%
\bibitem [{\citenamefont {Wang}\ \emph {et~al.}(2016)\citenamefont {Wang},
  \citenamefont {Qureshi}, \citenamefont {Yasin}, \citenamefont {Mukhin},
  \citenamefont {Ressouche}, \citenamefont {Zherlitsyn}, \citenamefont
  {Skourski}, \citenamefont {Geshev}, \citenamefont {Ivanov}, \citenamefont
  {Gospodinov},\ and\ \citenamefont {Skumryev}}]{wang_magnetoelectric_2016}%
  \BibitemOpen
  \bibfield  {author} {\bibinfo {author} {\bibfnamefont {Z.}~\bibnamefont
  {Wang}}, \bibinfo {author} {\bibfnamefont {N.}~\bibnamefont {Qureshi}},
  \bibinfo {author} {\bibfnamefont {S.}~\bibnamefont {Yasin}}, \bibinfo
  {author} {\bibfnamefont {A.}~\bibnamefont {Mukhin}}, \bibinfo {author}
  {\bibfnamefont {E.}~\bibnamefont {Ressouche}}, \bibinfo {author}
  {\bibfnamefont {S.}~\bibnamefont {Zherlitsyn}}, \bibinfo {author}
  {\bibfnamefont {Y.}~\bibnamefont {Skourski}}, \bibinfo {author}
  {\bibfnamefont {J.}~\bibnamefont {Geshev}}, \bibinfo {author} {\bibfnamefont
  {V.}~\bibnamefont {Ivanov}}, \bibinfo {author} {\bibfnamefont
  {M.}~\bibnamefont {Gospodinov}},\ and\ \bibinfo {author} {\bibfnamefont
  {V.}~\bibnamefont {Skumryev}},\ }\href {https://doi.org/10.1038/ncomms10295}
  {\bibfield  {journal} {\bibinfo  {journal} {Nat. Comm.}\ }\textbf {\bibinfo
  {volume} {7}},\ \bibinfo {pages} {10295} (\bibinfo {year}
  {2016})}\BibitemShut {NoStop}%
\bibitem [{\citenamefont {Lelièvre-Berna}\ \emph
  {et~al.}(2005{\natexlab{a}})\citenamefont {Lelièvre-Berna}, \citenamefont
  {Bourgeat-Lami}, \citenamefont {Gibert}, \citenamefont {Kernavanois},
  \citenamefont {Locatelli}, \citenamefont {Mary}, \citenamefont {Pastrello},
  \citenamefont {Petukhov}, \citenamefont {Pujol}, \citenamefont {Rouques},
  \citenamefont {Thomas}, \citenamefont {Thomas},\ and\ \citenamefont
  {Tasset}}]{LELIEVREBERNA2005141}%
  \BibitemOpen
  \bibfield  {author} {\bibinfo {author} {\bibfnamefont {E.}~\bibnamefont
  {Lelièvre-Berna}}, \bibinfo {author} {\bibfnamefont {E.}~\bibnamefont
  {Bourgeat-Lami}}, \bibinfo {author} {\bibfnamefont {Y.}~\bibnamefont
  {Gibert}}, \bibinfo {author} {\bibfnamefont {N.}~\bibnamefont {Kernavanois}},
  \bibinfo {author} {\bibfnamefont {J.}~\bibnamefont {Locatelli}}, \bibinfo
  {author} {\bibfnamefont {T.}~\bibnamefont {Mary}}, \bibinfo {author}
  {\bibfnamefont {G.}~\bibnamefont {Pastrello}}, \bibinfo {author}
  {\bibfnamefont {A.}~\bibnamefont {Petukhov}}, \bibinfo {author}
  {\bibfnamefont {S.}~\bibnamefont {Pujol}}, \bibinfo {author} {\bibfnamefont
  {R.}~\bibnamefont {Rouques}}, \bibinfo {author} {\bibfnamefont
  {F.}~\bibnamefont {Thomas}}, \bibinfo {author} {\bibfnamefont
  {M.}~\bibnamefont {Thomas}},\ and\ \bibinfo {author} {\bibfnamefont
  {F.}~\bibnamefont {Tasset}},\ }\href
  {https://doi.org/https://doi.org/10.1016/j.physb.2004.10.065} {\bibfield
  {journal} {\bibinfo  {journal} {Physica B: Condensed Matter}\ }\textbf
  {\bibinfo {volume} {356}},\ \bibinfo {pages} {141} (\bibinfo {year}
  {2005}{\natexlab{a}})},\ \bibinfo {note} {proceedings of the Fifth
  International Workshop on Polarised Neutrons in Condensed Matter
  Investigations}\BibitemShut {NoStop}%
\bibitem [{\citenamefont {B\"{o}ni}\ and\ \citenamefont
  {Keller}(1996)}]{Boni_1996_TASP}%
  \BibitemOpen
  \bibfield  {author} {\bibinfo {author} {\bibfnamefont {P.}~\bibnamefont
  {B\"{o}ni}}\ and\ \bibinfo {author} {\bibfnamefont {P.}~\bibnamefont
  {Keller}},\ }\href@noop {} {\bibfield  {journal} {\bibinfo  {journal}
  {PSI-Proceedings}\ }\textbf {\bibinfo {volume} {96-02}},\ \bibinfo {pages}
  {35} (\bibinfo {year} {1996})},\ \bibinfo {note} {proceedings of the 4th
  Summer School on Neutron Scattering, Zuoz, Switzerland, August 18-24,
  1996}\BibitemShut {NoStop}%
\bibitem [{\citenamefont {Tasset}(1989)}]{Tasset_1989_SNP}%
  \BibitemOpen
  \bibfield  {author} {\bibinfo {author} {\bibfnamefont {F.}~\bibnamefont
  {Tasset}},\ }\href
  {https://doi.org/https://doi.org/10.1016/0921-4526(89)90749-7} {\bibfield
  {journal} {\bibinfo  {journal} {Physica B: Condensed Matter}\ }\textbf
  {\bibinfo {volume} {156-157}},\ \bibinfo {pages} {627} (\bibinfo {year}
  {1989})}\BibitemShut {NoStop}%
\bibitem [{\citenamefont {Lelièvre-Berna}\ \emph
  {et~al.}(2005{\natexlab{b}})\citenamefont {Lelièvre-Berna}, \citenamefont
  {Bourgeat-Lami}, \citenamefont {Fouilloux}, \citenamefont {Geffray},
  \citenamefont {Gibert}, \citenamefont {Kakurai}, \citenamefont {Kernavanois},
  \citenamefont {Longuet}, \citenamefont {Mantegezza}, \citenamefont
  {Nakamura}, \citenamefont {Pujol}, \citenamefont {Regnault}, \citenamefont
  {Tasset}, \citenamefont {Takeda}, \citenamefont {Thomas},\ and\ \citenamefont
  {Tonon}}]{Lelievreberna_2005_SNP}%
  \BibitemOpen
  \bibfield  {author} {\bibinfo {author} {\bibfnamefont {E.}~\bibnamefont
  {Lelièvre-Berna}}, \bibinfo {author} {\bibfnamefont {E.}~\bibnamefont
  {Bourgeat-Lami}}, \bibinfo {author} {\bibfnamefont {P.}~\bibnamefont
  {Fouilloux}}, \bibinfo {author} {\bibfnamefont {B.}~\bibnamefont {Geffray}},
  \bibinfo {author} {\bibfnamefont {Y.}~\bibnamefont {Gibert}}, \bibinfo
  {author} {\bibfnamefont {K.}~\bibnamefont {Kakurai}}, \bibinfo {author}
  {\bibfnamefont {N.}~\bibnamefont {Kernavanois}}, \bibinfo {author}
  {\bibfnamefont {B.}~\bibnamefont {Longuet}}, \bibinfo {author} {\bibfnamefont
  {F.}~\bibnamefont {Mantegezza}}, \bibinfo {author} {\bibfnamefont
  {M.}~\bibnamefont {Nakamura}}, \bibinfo {author} {\bibfnamefont
  {S.}~\bibnamefont {Pujol}}, \bibinfo {author} {\bibfnamefont {L.-P.}\
  \bibnamefont {Regnault}}, \bibinfo {author} {\bibfnamefont {F.}~\bibnamefont
  {Tasset}}, \bibinfo {author} {\bibfnamefont {M.}~\bibnamefont {Takeda}},
  \bibinfo {author} {\bibfnamefont {M.}~\bibnamefont {Thomas}},\ and\ \bibinfo
  {author} {\bibfnamefont {X.}~\bibnamefont {Tonon}},\ }\href
  {https://doi.org/https://doi.org/10.1016/j.physb.2004.10.063} {\bibfield
  {journal} {\bibinfo  {journal} {Physica B: Condensed Matter}\ }\textbf
  {\bibinfo {volume} {356}},\ \bibinfo {pages} {131} (\bibinfo {year}
  {2005}{\natexlab{b}})},\ \bibinfo {note} {proceedings of the Fifth
  International Workshop on Polarised Neutrons in Condensed Matter
  Investigations}\BibitemShut {NoStop}%
\bibitem [{\citenamefont {Janoschek}\ \emph {et~al.}(2007)\citenamefont
  {Janoschek}, \citenamefont {Klimko}, \citenamefont {Gähler}, \citenamefont
  {Roessli},\ and\ \citenamefont {Böni}}]{Janoschek_2007_SNP}%
  \BibitemOpen
  \bibfield  {author} {\bibinfo {author} {\bibfnamefont {M.}~\bibnamefont
  {Janoschek}}, \bibinfo {author} {\bibfnamefont {S.}~\bibnamefont {Klimko}},
  \bibinfo {author} {\bibfnamefont {R.}~\bibnamefont {Gähler}}, \bibinfo
  {author} {\bibfnamefont {B.}~\bibnamefont {Roessli}},\ and\ \bibinfo {author}
  {\bibfnamefont {P.}~\bibnamefont {Böni}},\ }\href
  {https://doi.org/https://doi.org/10.1016/j.physb.2007.02.074} {\bibfield
  {journal} {\bibinfo  {journal} {Physica B: Condensed Matter}\ }\textbf
  {\bibinfo {volume} {397}},\ \bibinfo {pages} {125} (\bibinfo {year}
  {2007})}\BibitemShut {NoStop}%
\bibitem [{\citenamefont {Brown}\ \emph {et~al.}(1991)\citenamefont {Brown},
  \citenamefont {Chattopadhyay}, \citenamefont {Forsyth},\ and\ \citenamefont
  {Nunez}}]{Brown_1991_CuO_SNP}%
  \BibitemOpen
  \bibfield  {author} {\bibinfo {author} {\bibfnamefont {P.~J.}\ \bibnamefont
  {Brown}}, \bibinfo {author} {\bibfnamefont {T.}~\bibnamefont
  {Chattopadhyay}}, \bibinfo {author} {\bibfnamefont {J.~B.}\ \bibnamefont
  {Forsyth}},\ and\ \bibinfo {author} {\bibfnamefont {V.}~\bibnamefont
  {Nunez}},\ }\href {https://doi.org/10.1088/0953-8984/3/23/016} {\bibfield
  {journal} {\bibinfo  {journal} {J. of Phys.: Condensed Matter}\ }\textbf
  {\bibinfo {volume} {3}},\ \bibinfo {pages} {4281} (\bibinfo {year}
  {1991})}\BibitemShut {NoStop}%
\bibitem [{\citenamefont {Ain}\ \emph {et~al.}(1992)\citenamefont {Ain},
  \citenamefont {Menelle}, \citenamefont {Wanklyn},\ and\ \citenamefont
  {Bertaut}}]{Ain_1992_CuO_SNP}%
  \BibitemOpen
  \bibfield  {author} {\bibinfo {author} {\bibfnamefont {M.}~\bibnamefont
  {Ain}}, \bibinfo {author} {\bibfnamefont {A.}~\bibnamefont {Menelle}},
  \bibinfo {author} {\bibfnamefont {B.~M.}\ \bibnamefont {Wanklyn}},\ and\
  \bibinfo {author} {\bibfnamefont {E.~F.}\ \bibnamefont {Bertaut}},\ }\href
  {https://doi.org/10.1088/0953-8984/4/23/009} {\bibfield  {journal} {\bibinfo
  {journal} {J. Phys.: Condens. Matter}\ }\textbf {\bibinfo {volume} {4}},\
  \bibinfo {pages} {5327} (\bibinfo {year} {1992})}\BibitemShut {NoStop}%
\bibitem [{\citenamefont {Babkevich}\ \emph {et~al.}(2012)\citenamefont
  {Babkevich}, \citenamefont {Poole}, \citenamefont {Johnson}, \citenamefont
  {Roessli}, \citenamefont {Prabhakaran},\ and\ \citenamefont
  {Boothroyd}}]{Babkevich_2012_CuO_SNP}%
  \BibitemOpen
  \bibfield  {author} {\bibinfo {author} {\bibfnamefont {P.}~\bibnamefont
  {Babkevich}}, \bibinfo {author} {\bibfnamefont {A.}~\bibnamefont {Poole}},
  \bibinfo {author} {\bibfnamefont {R.~D.}\ \bibnamefont {Johnson}}, \bibinfo
  {author} {\bibfnamefont {B.}~\bibnamefont {Roessli}}, \bibinfo {author}
  {\bibfnamefont {D.}~\bibnamefont {Prabhakaran}},\ and\ \bibinfo {author}
  {\bibfnamefont {A.~T.}\ \bibnamefont {Boothroyd}},\ }\href
  {https://doi.org/10.1103/PhysRevB.85.134428} {\bibfield  {journal} {\bibinfo
  {journal} {Phys. Rev. B}\ }\textbf {\bibinfo {volume} {85}},\ \bibinfo
  {pages} {134428} (\bibinfo {year} {2012})}\BibitemShut {NoStop}%
\bibitem [{\citenamefont {Ouladdiaf}\ \emph {et~al.}(2006)\citenamefont
  {Ouladdiaf}, \citenamefont {Archer}, \citenamefont {McIntyre}, \citenamefont
  {Hewat}, \citenamefont {Brau},\ and\ \citenamefont
  {York}}]{Ouladdiaf_2006_OrientExpress}%
  \BibitemOpen
  \bibfield  {author} {\bibinfo {author} {\bibfnamefont {B.}~\bibnamefont
  {Ouladdiaf}}, \bibinfo {author} {\bibfnamefont {J.}~\bibnamefont {Archer}},
  \bibinfo {author} {\bibfnamefont {G.}~\bibnamefont {McIntyre}}, \bibinfo
  {author} {\bibfnamefont {A.}~\bibnamefont {Hewat}}, \bibinfo {author}
  {\bibfnamefont {D.}~\bibnamefont {Brau}},\ and\ \bibinfo {author}
  {\bibfnamefont {S.}~\bibnamefont {York}},\ }\href
  {https://doi.org/https://doi.org/10.1016/j.physb.2006.05.337} {\bibfield
  {journal} {\bibinfo  {journal} {Physica B: Condensed Matter}\ }\textbf
  {\bibinfo {volume} {385-386}},\ \bibinfo {pages} {1052} (\bibinfo {year}
  {2006})}\BibitemShut {NoStop}%
\bibitem [{\citenamefont {Soh}\ \emph {et~al.}(2021)\citenamefont {Soh},
  \citenamefont {Boothroyd}, \citenamefont {Prabhakaran}, \citenamefont
  {Qureshi}, \citenamefont {Rodr\'{i}guez-Velamaz\'{a}n}, \citenamefont
  {Rønnow}, \citenamefont {Spaldin},\ and\ \citenamefont
  {Stunault}}]{ILL_data}%
  \BibitemOpen
  \bibfield  {author} {\bibinfo {author} {\bibfnamefont {J.-R.}\ \bibnamefont
  {Soh}}, \bibinfo {author} {\bibfnamefont {A.~T.}\ \bibnamefont {Boothroyd}},
  \bibinfo {author} {\bibfnamefont {D.}~\bibnamefont {Prabhakaran}}, \bibinfo
  {author} {\bibfnamefont {N.}~\bibnamefont {Qureshi}}, \bibinfo {author}
  {\bibfnamefont {A.~J.}\ \bibnamefont {Rodr\'{i}guez-Velamaz\'{a}n}}, \bibinfo
  {author} {\bibfnamefont {H.~M.}\ \bibnamefont {Rønnow}}, \bibinfo {author}
  {\bibfnamefont {N.}~\bibnamefont {Spaldin}},\ and\ \bibinfo {author}
  {\bibfnamefont {A.}~\bibnamefont {Stunault}},\ }\bibfield  {journal}
  {\bibinfo  {journal} {Institut Laue-Langevin (ILL)}\ }\href
  {https://doi.org/10.5291/ILL-DATA.5-41-1172} {10.5291/ILL-DATA.5-41-1172}
  (\bibinfo {year} {2021})\BibitemShut {NoStop}%
\bibitem [{Bra()}]{Bransden}%
  \BibitemOpen
  \href@noop {} {\bibinfo {title} {See, for instance, \uppercase{B}.
  \uppercase{H}. \uppercase{B}ransden and \uppercase{C}. \uppercase{J}.
  \uppercase{J}oachain, \uppercase{P}hysics of \uppercase{A}toms and
  \uppercase{M}olecules, 2nd ed. (\uppercase{P}rentice \uppercase{H}all,
  \uppercase{E}nglewood \uppercase{C}liffs, \uppercase{NJ}, 2003), p.
  96}}\BibitemShut {NoStop}%
\end{thebibliography}%

\end{document}